\documentclass[aps,prx,twocolumn,groupedaddress]{revtex4-1}

\usepackage{xcolor}

\usepackage{amsmath}
\usepackage{amssymb}
\usepackage[hidelinks]{hyperref}

\newcommand{\be}{\begin{eqnarray}}
\newcommand{\ee}{\end{eqnarray}}
\newcommand{\ket}[1]{\ensuremath{\left| {#1} \right>}}
\newcommand{\bra}[1]{\ensuremath{\left< {#1} \right|}}

\newcommand{\Exp}[1]{\mathrm{e}^{#1}}%					Exponent
\newcommand{\micron}{\textmu m}

\usepackage{latexsym}
\usepackage{graphics}
\usepackage{multirow}
\usepackage{textcomp}

\usepackage{booktabs}
\usepackage{longtable}
\usepackage{float}
\usepackage{array}

\usepackage{pgfplots}
\usepackage{amssymb}

  \pgfplotsset{compat=newest}
  \usetikzlibrary{plotmarks}
  \usetikzlibrary{arrows.meta}
  \usepgfplotslibrary{patchplots}
  \usepackage{grffile}
  \usepgfplotslibrary{external}
\begin{document}
\title{Scalable arrays of micro-Penning traps for quantum computing and simulation}

\author{S.~Jain, J.~Alonso, M.~Grau and J.~P.~Home  }
%Star should be commented - these authors contributed equally to this work.

\affiliation{Institute for Quantum Electronics, ETH Z\"urich, Otto-Stern-Weg 1, 8093 Z\"urich, Switzerland}

%\ead{jhome@phys.ethz.ch}

\begin{abstract}
We propose the use of 2-dimensional Penning trap arrays as a scalable platform for quantum simulation and quantum computing with trapped atomic ions. This approach involves placing arrays of micro-structured electrodes defining static electric quadrupole sites in a magnetic field, with single ions trapped at each site and coupled to neighbors via the Coulomb interaction. We solve for the normal modes of ion motion in such arrays, and derive a generalized multi-ion invariance theorem for stable motion even in the presence of trap imperfections. We use these techniques to investigate the feasibility of quantum simulation and quantum computation in fixed ion lattices. In homogeneous arrays, we show that sufficiently dense arrays are achievable, with axial, magnetron and cyclotron motions exhibiting inter-ion dipolar coupling with rates significantly higher than expected decoherence. With the addition of laser fields these can realize tunable-range interacting spin Hamiltonians. We also show how local control of potentials allows isolation of small numbers of ions in a fixed array and can be used to implement high fidelity gates. The use of static trapping fields means that our approach is not limited by power requirements as system size increases, removing a major challenge for scaling which is present in standard radio-frequency traps. Thus the architecture and methods provided here appear to open a path for trapped-ion quantum computing to reach fault-tolerant scale devices.
\end{abstract}

\pacs{pacs}
\maketitle

The study of many-body physics in quantum mechanics is hindered by the inability of classical computing devices to store and manipulate the information required to specify these systems beyond about 50 spins \cite{82Feynman,17Haner}. Quantum devices, which directly work in Hilbert space, would overcome these limitations and furthermore open up access to other calculations beyond the reach of classical supercomputers  \cite{82Feynman,14Georgescu,13Schaetz,12Blatt}.  Trapped atomic ions are among the most successful platforms for exploring these advances. They interact via the long-range Coulomb force, which when combined with laser and microwave fields has allowed high quality quantum logic gates \cite{16Ballance, Gaebler2016} as well as Hamiltonian engineering for quantum simulations \cite{17Zhang}. The most successful approach to controlling multiple trapped-ion qubits has involved the use of semi-rigid ion crystals, formed through the competing energy requirements of the global trapping potential and the Coulomb repulsion. Primarily these use 1-dimensional ion chains in radio-frequency (r.f.) ion traps \cite{11Monz, 17Zhang}. The restriction to one dimension is imposed by the desire to trap at the null of the r.f.~field. Penning traps have been used to perform quantum simulations with 2-dimensional (2-d) ion crystals, with the complication of continuous rotation in the crystal plane \cite{12Britton}. In both cases, the intrinsic link between the lattice structure and the oscillation frequencies of the normal modes of oscillation places constraints which limit the range of physics which can be investigated. Furthermore, neither of these approaches is well-suited to scaling these systems up to levels close to a million qubits which are expected to be required for solving relevant problems in quantum chemistry \cite{Wecker2014}.

An alternative platform which increases flexibility is the use of micro-fabricated ion traps with electrode structures on length scales close to the inter-ion separation. This would allow access to arbitrary 2-d lattice geometries, as well as providing the possibility to locally tune potentials in order to decouple subsets of ions to facilitate local multi-qubit gates. For r.f.~traps, early experiments in this direction have been carried out \cite{Wilson2014, 16Mielenz}, but a number of significant challenges to scaling arise from the use of radio-frequency potentials. One is that r.f.~power dissipation in the electrodes increases with the number of sites (similar to the challenge in scaling optical power which limits optical dipole traps for neutral atoms). It also appears difficult to achieve small site spacings for strong Coulomb coupling in extended arrays of ions  \cite{09Schmied,11Schmied,14Krauth}. A further practical challenge is that the operation of such traps relies on the alignment of the microscopic static and r.f.~quadrupole potentials, which is challenging to achieve in the presence of stray charges on the electrode surfaces, especially when large numbers of sites are involved \cite{98Berkeland2}.

Surface electrode Penning traps offer an alternative platform for trapping ions which utilizes only \emph{static} fields. Surface-electrode Penning traps have been experimentally realized with both electrons \cite{10Goldman} and atomic ions \cite{Crick2010}, with the light mass of the former offering the potential for strong site-site couplings \cite{03Ciaramicoli,05Stahl, 10Goldman}. For electrons, arrays of such traps have been proposed as a coupled system suitable for quantum information \cite{05Stahl}. However electrons lack many of the control techniques available to atomic ions, which have the considerable advantage of being able to utilize advanced laser techniques for cooling, initialization, detection and control \cite{08Bushev, Hellwig2010}.

In this Letter, we consider the use of arrays of surface Penning traps to realize 2-d lattices of trapped atomic ions for quantum simulation and computation, a setting which appears scalable due to the use of only static fields for trapping. We present a detailed study of the collective motional modes of oscillations for ions trapped in this architecture, which is the crucial element for any quantum information application. We find that the motion of $N$ coupled ions can be mapped onto the same quadratic eigenvalue problem for both the classical and quantum treatments, and use this framework to generalize the single-ion invariance theorem for stable modes of motion \cite{82Brown,86Brown} to an arbitrary number of ions. We then focus on two cases relevant to quantum simulation and computation in extensible ion lattices. For quantum simulation, we demonstrate that closely spaced arrays can be produced, by obtaining optimal electrode geometries for a range of different lattices. Here Penning traps allow significantly higher ion densities and thus higher coupling than radio-frequency traps for similar experimental constraints. We show that variable range spin-spin interactions can be produced, as well as verifying the ability to access useful ion temperatures through laser cooling. We follow this by studying the feasibility of quantum computing using multi-qubit gates on two nearest-neighbor ions which are embedded in an array, by using local mode decoupling of the ions. We show that a special regime of large zero-point motion available for the magnetron and cyclotron modes allows fast high fidelity logic gates, which could be applied to lattice-based error-correction codes such as the surface and topological color codes \cite{Fowler2012, Bombin2006}.

\section{Penning microtrap array}
\begin{figure}
  \centering
  \begin{tabular}{c}
   \LARGE
  \resizebox{\columnwidth}{!}{\includegraphics{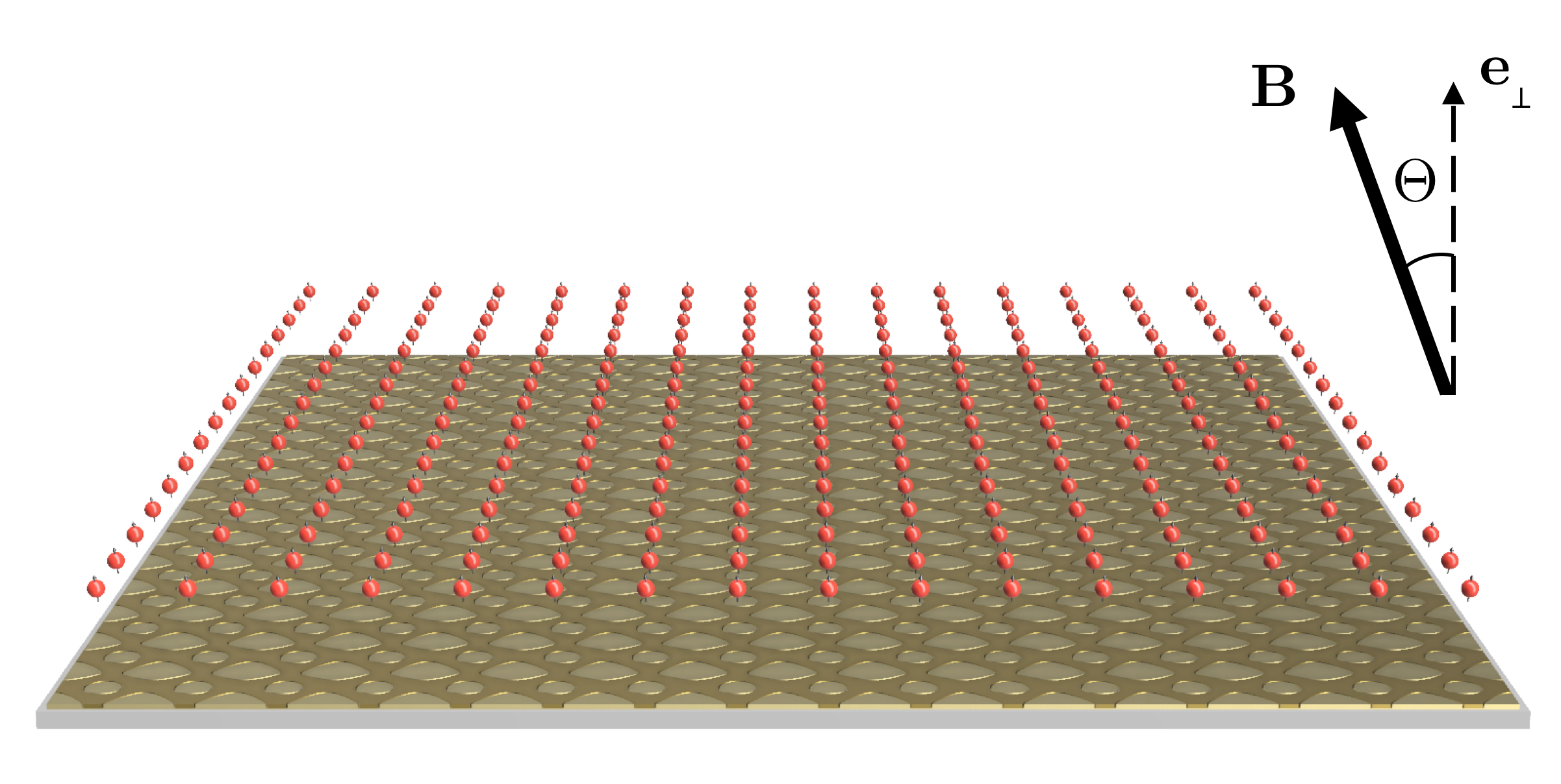}}\\
  i.
  \\
  \resizebox{0.6\columnwidth}{!}{\includegraphics{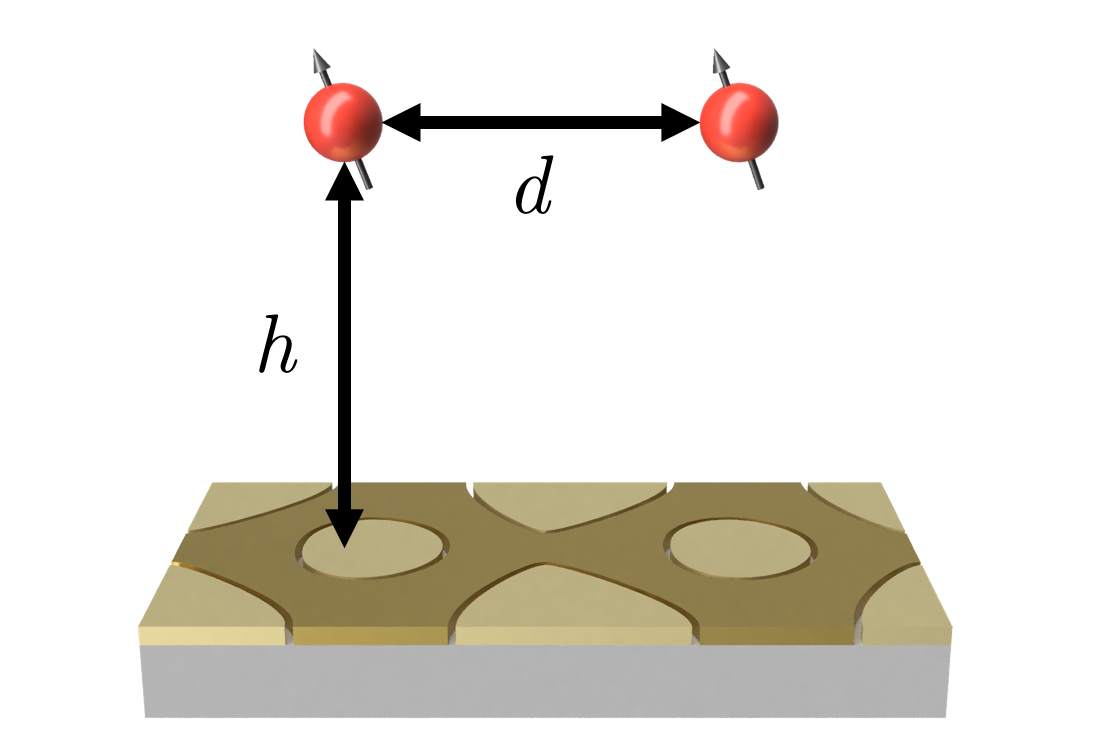}}\\
  ii.
   \end{tabular}
 \caption{Proposed array of Penning microtraps. The magnetic field $\mathbf{B}$ makes an angle $\Theta$ with $\mathbf{e}_{\perp}$, the vector normal to the lattice plane in which all of the ions lie. The ion separation is given by the electrode pattern. The critical parameter for the ion-ion interaction is the separation of neighboring sites $d$, while the distance from the surface $h$ strongly affects the level of noise from fluctuating electric fields \cite{00Turchette}.}
\label{fig:array}
\end{figure}

The configuration that we propose is shown in figure \ref{fig:array}. Electric potentials are applied to trap electrodes laid out on a planar structure such as to form an array of static electric 3-d quadrupoles at a distance $h$ above the trap surface. These, combined with the homogeneous magnetic field, provide three-dimensional confinement of a single ion at each site.  Together with the trapping potential the Coulomb interaction defines the equilibrium positions of the ions. For an infinite lattice, the equilibrium positions will align with the centers of the quadrupole potentials created through the periodic arrangement of electrodes. For the finite case, this is not true and should be taken into account while designing the electrodes to generate the required lattice of equilibrium positions. The vibrations of the ions around their equilibrium positions is also coupled by the Coulomb interaction. As we will show in the next section, at the single quantum level this depends as the inverse cube on the distance between ions and the zero-point motion amplitude, and has the strongest effect for two resonant sites. From the trapping perspective, the primary objective is to achieve closely spaced ions. The desirable electrode geometries for infinite lattices will be considered in more detail in section \ref{sec:Qsim}, and then applied to quantum simulation problems. A secondary aspect is that the electrodes are microscopic, which allows local control. In this way, ions may be tuned in and out of resonance with the rest. This will be studied in the context of quantum computing in section \ref{sec:Qcomp}. In section \ref{sec:example-implementation} we present a realistic scenario in which these ideas could be implemented in an experiment.

\section{Dipolar ion-ion coupling}

The basis for multi-qubit quantum control and quantum simulations in such a setting is the long-range Coulomb interaction, which couples the oscillations of individual ions at different sites. While for an ion oscillating along a single spatial axis it is simple to think of the oscillating charge distribution as an oscillating dipole potential, and thus that the ions are coupled through something similar to a dipole-dipole coupling, the types of motion exhibited by ions in a Penning trap are more complex. To get a feeling for the nature of this coupling, we first consider for the moment a simplified setting, in which each ion is trapped in a cylindrically symmetric static quadrupole potential $m \omega_z^2/2 (z^2 - (x^2 + y^2)/2)$ for ions of mass $m$ and charge $e$ embedded in a magnetic field of strength $B_0$ aligned along the $z$ axis.

%At a single site, the potential and magnetic field give rise to a Hamiltonian
%\begin{equation}
%\begin{aligned}
%\hat{H}_s &= \frac{\hat{p}^2_x + \hat{p}^2_y}{2 m} + \frac{1}{8} m \omega^2_1 \left( \hat{x}^2 + \hat{y}^2 \right) - \frac{\omega_c}{2}  \left( \hat{x} \hat{p}_y - \hat{y} \hat{p}_x \right) \\
%&+ \frac{\hat{p}^2_z}{2 m} + \frac{1}{2} m \omega^2_z \hat{z}^2
%\end{aligned}
%\end{equation}
%where $\omega_1 = \sqrt{\omega_c^2 - 2 \omega_z^2}$ and $\omega_c = e B_0/m$ is the bare cyclotron frequency. Writing the position and momentum operators in terms of creation and annihilation operators for the individual $x$, $y$, and $z$ degrees of freedom, defined as
%\begin{equation}
%\begin{aligned}
%\hat{x} = \sqrt{\frac{ \hbar}{m \omega_1}} \left( \hat{a}^{\dagger}_x + \hat{a}_x \right) &, \quad \hat{p}_x = i \sqrt{\frac{\hbar m \omega_1}{4}} \left( \hat{a}^{\dagger}_x - \hat{a}_x \right), \\
%\hat{y} = \sqrt{\frac{ \hbar}{m \omega_1}} \left( \hat{a}^{\dagger}_y + \hat{a}_y \right) &, \quad \hat{p}_y = i \sqrt{\frac{\hbar m \omega_1}{4}} \left( \hat{a}^{\dagger}_y - \hat{a}_y \right), \\
%\hat{z}  = \sqrt{\frac{\hbar}{2 m \omega_z}} \left( \hat{a}^{\dagger}_z + \hat{a}_z \right) &, \quad \hat{p}_z = i \sqrt{\frac{\hbar m \omega_z}{2}} \left( \hat{a}^{\dagger}_z - \hat{a}_z \right), \\
%\end{aligned}
%\end{equation}
At a single site, the Hamiltonian for an ion in such a potential can be written as
%\be
%\hat{H}_s &=&
%\frac{\hbar \omega_1}{2} \left(\hat{a}_x^\dagger \hat{a}_x+\hat{a}_y^\dagger \hat{a}_y + 1\right) + i \frac{\hbar \omega_c}{2}\left(\hat{a}_x^\dagger \hat{a}_y - \hat{a}_y^\dagger \hat{a}_x \right) \nonumber \\ && + \hbar \omega_z \left(\hat{a}_z^\dagger \hat{a}_z + 1/2\right) .
%\ee
%
%This can be separated into a sum of three independent harmonic oscillators using the transformation $
%\hat{a}_{\pm} = \frac{1}{\sqrt{2}} \left( \hat{a}_x \pm i \hat{a}_y \right)$
%for the radial motion. We then obtain
\begin{equation}
\begin{aligned}
\label{eq:HsinglePenning}
\hat{H}_s &=  \hbar \omega_+ \left(\hat{a}^\dagger_+ \hat{a}_+ + 1/2\right) - \hbar \omega_- \left(\hat{a}^\dagger_- \hat{a}_-
+ 1/2\right) \\
&\hphantom{+}+ \hbar \omega_z \left(\hat{a}^\dagger_z \hat{a}_z + 1/2\right),
\end{aligned}
\end{equation}
where $\omega_{\pm} = (\omega_c \pm \omega_1)/2$ are the frequencies of the modified cyclotron motion and magnetron motion respectively, with $\omega_1 = \sqrt{\omega_c^2 - 2 \omega_z^2}$ and $\omega_c = e B_0/m$ the bare cyclotron frequency \cite{86Brown}. Here $\hat{a}_{\pm}$ and $\hat{a}_z$ are the annihilation operators for the corresponding modes. This Hamiltonian differs from a standard 3-dimensional oscillator due to the negative sign in front of the magnetron term. It is also worth noting that both the magnetron and cyclotron motions are two-dimensional, thus the relevant creation and annihilation operators are made up of sums of the form $\hat{a}_{\pm} = \frac{1}{\sqrt{2}} \left( \hat{a}_x \pm i \hat{a}_y \right)$ where $\hat{a}_x, \hat{a}_y$ refer to motion along $x$ and $y$ respectively (full details of the transformations and definitions used can be found in Appendix \ref{app:SingleSite}).

Let us now consider two such sites labelled by indexes $i$ and $j$ containing ions with equilibrium positions separated by the vector ${\bf R}_{ij,0} = R_{ij,0}(\sin(\theta_{ij})\cos(\phi_{ij}), \sin(\theta_{ij})\sin(\phi_{ij}),\cos(\theta_{ij})) $ which has magnitude $R_{ij,0}$ and makes an angle $\theta_{ij}$ with the magnetic field. For the current argument, let us work in the approximation that the motion of the ions can be assumed to be a small perturbation which is well-described using a second order expansion of the Coulomb interaction about the equilibrium positions. Moving to a rotating frame with respect to $\hat{H}_s$ for the operators at each of the sites and further assuming that the difference frequency between the bare modes is much larger than the respective exchange frequencies for the different modes, we find the Coulomb interaction Hamiltonian
\be
\hat{H}_{c,ij} &=&  \sum_{\nu} \hbar \Omega_{\nu, {\rm ex}}^{ij} K_{\nu} \left(\hat{a}_{\nu,i}^\dagger \hat{a}_{\nu,j}+\hat{a}_{\nu,j}^\dagger \hat{a}_{\nu,i}\right) \nonumber \\
&& - \sum_{\nu} \hbar \Omega_{\nu, {\rm ex}}^{ij} K_{\nu} \left(\hat{a}_{\nu,i}^\dagger \hat{a}_{\nu,i}+ \hat{a}_{\nu,i} \hat{a}_{\nu,i}^\dagger \right), \label{eq:dipoledipole}
\ee
where $K_z = -K_{\pm} =  1 - 3 \cos^2(\theta_{ij})$. The first term corresponds to hopping of excitations between the sites, while the second gives the modification of the on-site energy due to the static potential of the other ion. The respective coupling strengths for hopping of vibrational quanta from one ion to another are given by the exchange frequencies
\begin{equation}
\label{eq:Omex}
\Omega_{\nu, \rm ex}^{ij} = \frac{e^2}{4 \pi \epsilon_0 m \omega'_\nu R_{ij,0}^3},
\end{equation}
where $\omega'_z = \omega_z$ and $\omega'_{+}, \omega'_{-} = \omega_1$.

The coupling Hamiltonian has a dipolar form for all modes. The sign of the coupling for the magnetron and cyclotron modes is inverted with respect to that of the axial motion. For each type of motion the orientation of the effective dipole is along the magnetic field axis. While this is expected for the axial oscillation, for the other modes this is less intuitive, since it corresponds to a direction perpendicular to the plane of oscillation of both the cyclotron and magnetron motions. When the coupling above is generalized to a two-dimensional lattice of sites, the anisotropy of the interactions in the lattice plane depends on the angle $\Theta$ between the magnetic field and the lattice normal. For $\Theta = 0^\circ$ the interactions are isotropic because $\theta_{ij} = \pi/2$ for all directions within the plane. For $\Theta = 90^\circ$, $\theta_{ij}$ values vary between $0$ and $2\pi$, thus the interactions are anisotropic.

An additional feature of the couplings, which we make use of in section \ref{sec:Qcomp} is that the couplings for the modified cyclotron and magnetron motions are proportional to the reciprocal of $\omega_1$, rather than the mode frequency itself, a result which stems from the dependence of the zero-point motion of each of these modes. Thus by tuning $\omega_1$ to be small, the coupling between two sites can be enhanced. This can be performed by raising $\omega_z$ towards $\omega_c/\sqrt{2}$ (although the limit $\omega_z = \omega_{c} / \sqrt{2}$ is unstable). Unlike a standard mechanical oscillator, this enhancement of the zero-point motion does not require lowering of the mechanical oscillator frequency.

\section{Operating conditions}
\begin{figure}
  \centering
  \begin{tabular}{c}
    \Large
  \resizebox{0.7\columnwidth}{!}{\includegraphics{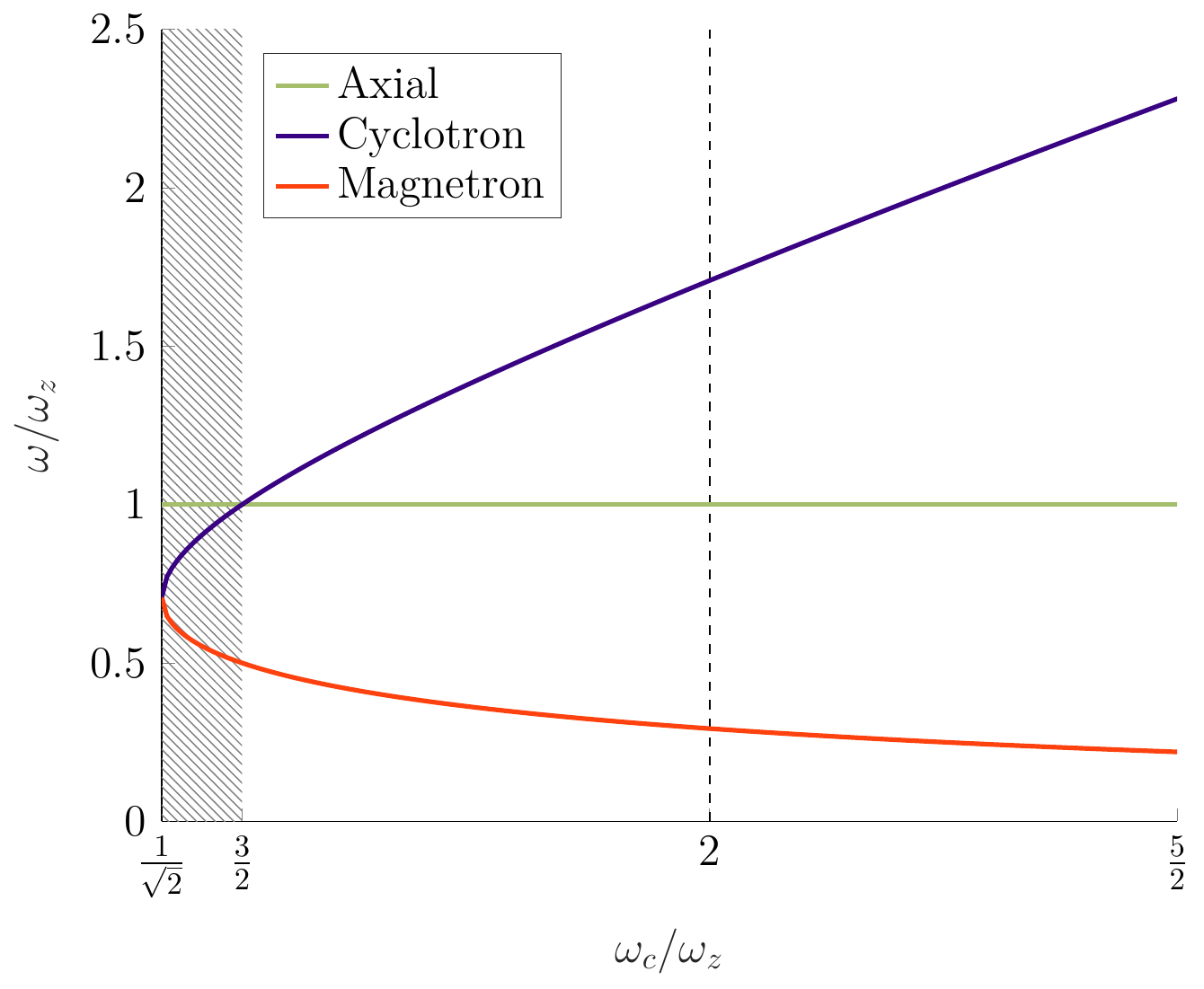}}
 \\
\end{tabular}
\caption{Frequencies of oscillation of a single ion Penning trap as a function of the bare cyclotron frequency. All frequencies are given in units of the axial frequency. }
\label{fig:singleionfreqs}
\end{figure}

Within the restrictions of using symmetric quadrupole potentials, there is nevertheless a range of possible choices of $\omega_c, \omega_z$ which could be used, related to the type of couplings which are desired and the achievable experimental constraints. However these choices are constrained by the fixed dependence of the mode frequencies on these two parameters. As an example which we find interesting from the perspective of engineered tunable-range spin-spin couplings (see section \ref{sec:spinspin}), we consider the challenge of operating a trap with a fixed achievable $\omega_z$ while desiring that the modes have a large enough splitting that three separated bands of modes occur corresponding to coupled magnetron, modified cyclotron and axial frequencies.  We find that for an infinite lattice the width of any one band is in the range between 3 and 6 times the exchange frequency for the respective mode. Figure \ref{fig:singleionfreqs} shows the trap frequencies which can be achieved normalized to the axial frequency $\omega_z$. To achieve large mode separations, it appears attractive to work in the regime for which $\omega_+ > \omega_z$, which corresponds to $\omega_c> 3 \omega_z/2$. However note that this implies that $\omega_- < \omega_z/2$. The splitting between the $\omega_\pm$ modes and the axial modes becomes equal in magnitude when $\omega_c = 2 \omega_z$ (see dashed line), for which $\omega_- = \omega_z(\sqrt{2}-1)/\sqrt{2} \approx 0.29 \omega_z$. In practice there are good reasons to work with high magnetron frequencies, for example because this would be expected to result in lower heating rates of those modes due to sampling noise at higher frequencies - this may limit the choice of trap parameters.

A second regime of operation could also be used, in which the difference in mode frequencies is significantly below the trap frequencies themselves. For this regime, the axial frequency can be chosen to be the highest frequency of the three, and the modified cyclotron and magnetron modes become closer in frequency. This can be achieved in a narrow regime for which $3 \omega_z/2 > \omega_c > \sqrt{2} \omega_z $, as indicated by the shaded area in figure \ref{fig:singleionfreqs}. As $\omega_c$ is reduced within this range, the difference frequency $\omega_1$ between the modified cyclotron and magnetron frequencies is reduced. This increases the exchange frequencies for these modes (but also proportionally the heating rate). The maximal separation of modes within this regime is $0.191 \omega_z$.

\section{Normal Mode Analysis}

A full analysis of the normal modes of vibration of ions is a prerequisite to understand the methods utilized for laser cooling and implementation of analog quantum simulation. The presence of a magnetic field in Penning traps makes this analysis non-trivial. While the case of a naturally formed two-dimensional ion crystal in a macroscopic Penning trap has been treated before, this involves a transformation to a frame co-rotating with the rigid ion crystal \cite{13Wang}. Here we avoid any such transformation, and derive the normal modes for the general case of ions in an array of micro-Penning traps, assuming only that the equilibrium positions are well defined so that the harmonic approximation for the electric potential can be used and that the magnetic field is uniform over the region explored by each ion. The analysis leads to a generalization of a well known invariance theorem used for single ions in Penning traps \cite{82Brown,86Brown}.

\subsection{Classical Treatment}
In the classical regime, the normal mode analysis of a finite system of $N$ trapped ions can be carried out with the help of Lagrangian mechanics. For simplicity, it is assumed that all ions have an identical mass $m$, but the analysis can be generalized to systems containing ions with different masses (see Appendix \ref{app:Classical}). In the frame of reference of the laboratory, with no oscillatory fields present, the Lagrangian of the system is given by
\begin{equation}
L = \sum^{N}_{j=1}\left[ \frac{1}{2}m|\dot{\mathbf{R}}_{j}|^{2}+e\mathbf{A}_j \cdot \dot{\mathbf{R}}_j-e\Phi_{j}\right],
\end{equation}
where $\mathbf{R}_{j}$ denotes the lab coordinates of ion $j$, $\mathbf{A}_j = \frac{1}{2}(\mathbf{B}\times\mathbf{R}_j)$ is the vector potential in the symmetric gauge due to the uniform magnetic field $\mathbf{B} = B_0 \hat{\bf e}_z$, and $\Phi_j$ is the electric potential containing a sum of contributions from the trapping potential for ion $j$ and the Coulomb interaction potential experienced by this ion due to all others.

The first step in obtaining normal modes is to solve for the equilibrium positions, which we carry out numerically. The second order term in the series expansion of the system Lagrangian about the equilibrium positions dictates the normal mode dynamics of the system near the stable spatial configuration. The equations of motion for the $3N$-dimensional vector $q=\left[ x_1 \, ...\, x_N \quad y_1 \,... \, y_N \quad z_1 \,... \, z_N \right]^T$ consisting of all the generalized position coordinates can then be deduced as
\begin{equation}
M\ddot{q}-W\dot{q}+\Phi q=0,
\end{equation}
where $M$, $W$ and $\Phi$ are $3N \times 3N$ matrices defined as $M = m \cdot \mathbb{I}_{3N}$, $W^{xy}_{jk} = - W^{yx}_{jk} = eB_0 \delta_{jk}$ and $\Phi^{\mu \nu}_{jk} = \partial q^{\mu}_{j} \partial q^{\nu}_{k} L$. $\Phi$ contains only terms from the static electric potential and the Coulomb interactions. Here, the indices $j, k$ run over the ion numbers $1$ to $N$ while the indices $\mu, \nu$ run over the Cartesian components $x, y, z$. The `mass matrix' $M$ is a real diagonal matrix, $W$ is a real antisymmetric matrix representing the velocity-dependent forces (often referred to as the `damping matrix'), while the `stiffness matrix' $\Phi$ is a real symmetric traceless matrix.

To find the normal modes of motion, we substitute the oscillating trial solution $q=q_0 \Exp{-i\omega t}$ which yields the Quadratic Eigenvalue Problem (QEP)
\begin{equation}
[\,\omega^2 M +\omega (-iW)-\Phi\, ]q_0=0,
\end{equation}
that can be solved for complex eigenvectors $q_0$ and eigenvalues $\omega$. When all eigenvalues are real the motion of all ions is bounded and stable confinement can be achieved. Each of the $3N$ collective normal modes of motion is thus characterized by the eigenpair $\{ \omega_{\lambda}, q_{\lambda} \}$ and the general solution for the motion can be expressed as
\begin{equation}
q(t) = \text{Re} \left[ \sum^{3N}_{\lambda=1} r_{\lambda} q_{\lambda} \Exp{-i \left( \omega_{\lambda} t + \delta_{\lambda} \right)} \right],
\end{equation}
with the amplitude $r_{\lambda}$ and phase $\delta_{\lambda}$ for each mode $\lambda$ determined by the initial conditions.

It is important to note that the total energy contained in each mode
\begin{equation}
\begin{aligned}
E_{\lambda} &=\frac{1}{4} r^2_{\lambda} \left( \omega^2_{\lambda}q^H_{\lambda}Mq_{\lambda}+q^H_{\lambda}\Phi q_{\lambda} \right)
\end{aligned}
\end{equation}
is not trivially positive, unlike the case of Paul traps. Typically we will observe $N$ modes dominated by motion along the axial direction and it is convenient to continue calling these modes axial modes in the context of the $N$ ion array of Penning traps. Each of the axial modes has a positive total mode energy. Similarly there are $2N$ radial modes out of which $N$ have each a positive mode energy and $N$ have each a negative mode energy. We will denote the radial modes with positive sign as cyclotron modes and the ones with negative sign as magnetron modes.

\subsection{Quantum Mechanical Treatment}
The solution for the normal modes in the quantum regime involves the formulation of the Hamiltonian in terms of the canonical position and momentum operators $\hat{q}_j$ and  $\hat{p}_j$, and then forming the phonon creation and annihilation operators, $\hat{a}^{\dagger}_\lambda$ and $\hat{a}_\lambda$, for each mode $\lambda$ as linear combinations of these operators,
\begin{equation}
\hat{a}^{\dagger}_\lambda=\sum^{3N}_{j=1}(\alpha_{\lambda j}\hat{p}_j+\beta_{\lambda j}\hat{q}_j),
\end{equation}
where $\alpha$ and $\beta$ are complex coefficients. The objective is to find these coefficients which allow us to diagonalize the Hamiltonian for a stable system in the second quantized form $\hat{H}=\sum^{3N}_{\lambda=1}\hbar\omega_{\lambda}(\hat{a}^{\dagger}_\lambda \hat{a}_\lambda + \frac{1}{2})$  with the phonon operators following the standard commutation relations,
$[\hat{a}^{\dagger}_{\lambda},\hat{a}^{\dagger}_{\lambda '}]=0$, $[\hat{a}_{\lambda},\hat{a}_{\lambda'}]=0$, $[\hat{a}_{\lambda},\hat{a}^{\dagger}_{\lambda'}]=\delta_{\lambda \lambda'}$. Combining these requirements, we find that the $3N$-dimensional vectors $\alpha_\lambda$ for each mode $\lambda$ satisfy the same QEP we had to solve in the classical analysis
\begin{equation}
[\omega^2 M + \omega (-iW) -\Phi]\alpha_\lambda =0,
\end{equation}
and we have the relation $\beta_\lambda=i\omega_\lambda M\alpha_\lambda+\frac{1}{2}W\alpha_\lambda$. These vectors can then be normalized such that the condition $[\hat{a}_{\lambda},\hat{a}^{\dagger}_{\lambda}]=1$ is fulfilled.

We note that the treatment discussed in this section encompasses both r.f.~and Penning traps, and the normal modes for the former under the pseudopotential approximation can be retrieved by choosing the magnetic field strength as zero and adding a suitable pseudopotential term to $\Phi$. In this case $\Phi$ is not trace zero.

\section{Generalized Invariance Theorem}

A real trap is imperfect and can suffer from misalignments between the magnetic field and the confining axis of the quadrupole potential, or the trap potential differing from the desired precise form.
Including these imperfections in the matrices $W$ and $\Phi$ in the QEP encountered in the classical analysis and dividing the equation by $m$, we get
\begin{equation}
[\omega^2 \cdot \mathbb{I}_{3N}+\omega (-iW^{\prime})-\Phi^{\prime} ]q_0=0,
\end{equation}
where we define the matrices $W^{\prime} = W /m $ and $\Phi^{\prime} = \Phi/m$. This new QEP can then be linearized by mapping it on to a standard eigenvalue problem while increasing the dimensionality by a factor of two so that we arrive at
\begin{equation}
A v = \omega v,
\end{equation}
with $6N$-dimensional eigenvectors $ v = \begin{bmatrix} q_0 & \omega q_0\end{bmatrix}^{T}$ and $6N$ eigenvalues $\omega$ belonging to the $6N \times 6N$ matrix $A$
\begin{equation}
A=\begin{bmatrix} \mathbb{O}_{3N} & \mathbb{I}_{3N} \\ \Phi^{\prime} & iW^{\prime} \end{bmatrix} .
\end{equation}
Since $ A^2 v =\omega^2 v $ and the sum of eigenvalues of a matrix is equal to its trace,
\begin{equation}
\begin{aligned}
\sum^{6N}_{\lambda=1}\omega^2_\lambda = \text{tr}(A^2)
= \text{tr}(2\Phi^{\prime}-W^{\prime 2})
=-\text{tr}(W^{\prime 2}) \ .
\end{aligned}
\end{equation}
This result will hold for any potential terms added to $\Phi'$ which exist in free space, since these must be traceless in order to satisfy Laplace's equation (it would not hold for a pseudopotential). While for a single ion only the trapping potentials are present, in the case of a multi-ion system the Coulomb interactions are also contained in this term. Noting that the frequencies come in pairs of positive-negative values in the stable regime we can express this sum in terms of the $3N$ positive frequencies,
\begin{equation}\label{eq:sum_m}
\begin{aligned}
\sum^{3N}_{\lambda=1}\omega^2_\lambda &=-\frac{1}{2}\text{tr}(W^{\prime 2}) \\
&=N\omega^2_c,
\end{aligned}
\end{equation}
where  $\omega_c = eB_0/m$  is the bare cyclotron frequency. This relation between the strength of the magnetic field quantified and the normal mode frequencies of a stable $N$-ion system represents a non-trivial generalization of the well known Brown-Gabrielse invariance theorem for a single ion in a Penning trap \cite{82Brown,86Brown},
\begin{equation}
\omega^2_+ + \omega^2_- +\omega^2_z = \omega^2_c.
\end{equation}
A complementary result relating the product of the normal mode frequencies to the curvature of the total electric potential can be derived by taking the determinant of the matrix $A$, giving

\begin{equation}\label{eq:prod_m}
\prod^{3N}_{\lambda=1} \left( m \omega^2_{\lambda} \right) = \lvert \Phi \rvert .
\end{equation}

The results (\ref{eq:sum_m}) and (\ref{eq:prod_m}) further generalize to systems containing ions of different masses (see Appendix \ref{sec:AppInvarianceMass} for more details). The single-ion invariance theorem is widely used in precision measurement \cite{09Gabrielse}, and our generalization can be applied to precision mass measurements of multi-ion crystals~\cite{19Gutierrez}.
% While we find our generalization to be of interest theoretically, it is unclear whether it has an application to precision measurement, since it seems challenging to measure the frequencies of a large number of modes.

\section{Laser Cooling}
Doppler cooling is more complicated in Penning traps as compared to radio-frequency traps due to the fact that the magnetron modes have a negative total energy. As a consequence the cooling requirements of the magnetron modes are incompatible with those of the axial and cyclotron modes, and no combination of uniform beams can cool all modes of motion simultaneously \cite{82Itano}. One way to combat this is to use a non-uniform beam with intensity gradient, but the final temperatures reached for both kinds of radial modes are greater than one would expect from the standard Doppler cooling limit, and the range of motional frequencies that allow for cooling all three kinds of modes is also restricted  \cite{82Itano}.

An alternative solution which has been realized experimentally \cite{02Powell} is to couple the cyclotron and magnetron modes by applying a weak quadrupolar electric field $\phi_{\rm ax} (x^2-y^2)$ oscillating at the bare cyclotron frequency, in a technique known as axialization \cite{00Thompson}. A red-detuned uniform-intensity Doppler cooling laser beam can then simultaneously cool all modes. With the axialization drive the system no longer consists solely of electrostatic fields but since the amplitude of such a drive is much lower than for the r.f.~drive required for Paul traps, the deleterious effects of micromotion are accordingly much smaller. Moreover, this technique is required only during the laser cooling process and works efficiently at all trap frequencies, allowing trapping in regimes not accessible through the use of just inhomogeneous beams.

\begin{figure}
  \centering
  \Large
    \resizebox{\columnwidth}{!}{\includegraphics{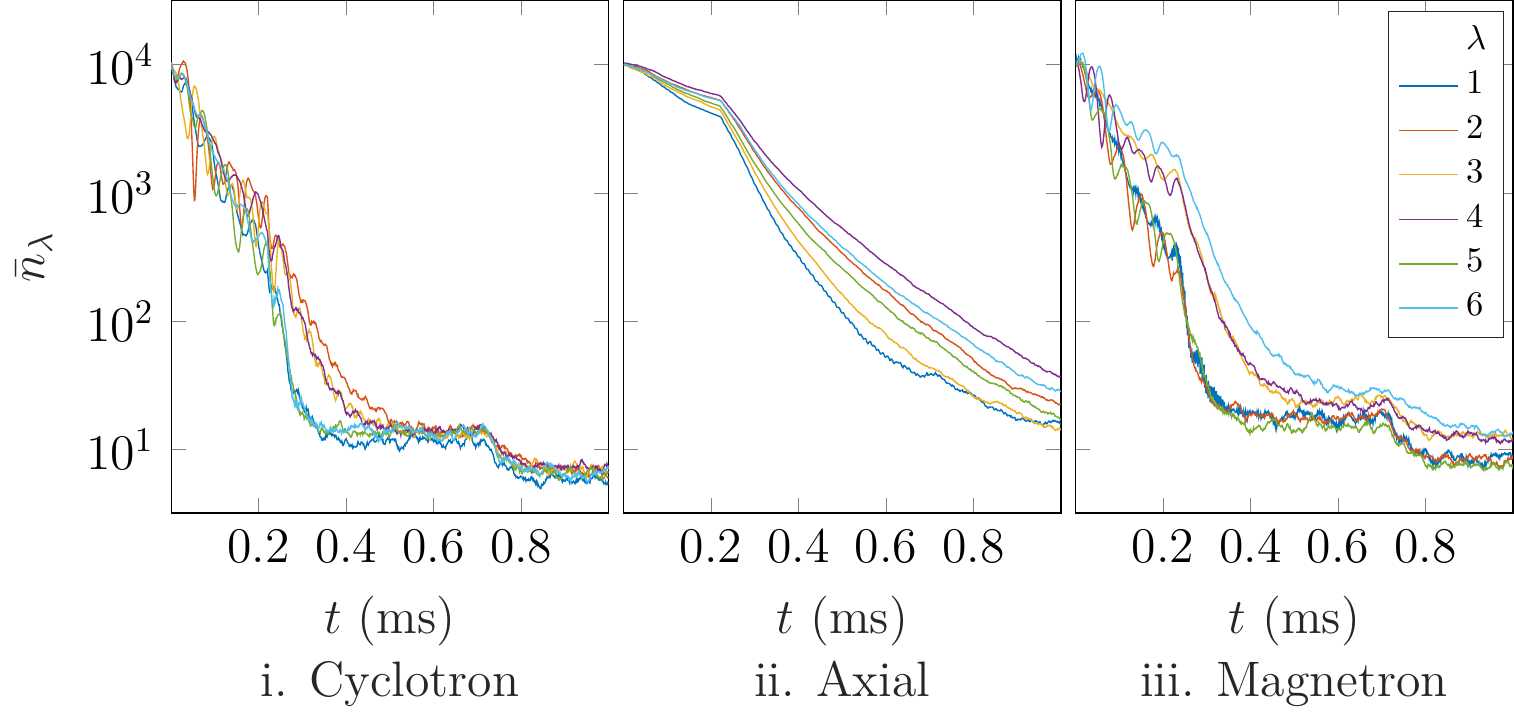}}
 \caption{Mode occupation numbers as a function of time for a six $^9$Be$^+$ ion honeycomb lattice being laser cooled in conjunction with axialization}  \label{fig:coolarray}
\end{figure}

The derivation of the rate equations of the mode amplitudes of ions due to Doppler cooling in Penning traps in the presence of axialization has no simple analytical solution. Instead, we perform numerical simulations on small numbers of ions in which we numerically integrate the equations of motion of the trapped ion including the axialization potential as well as stochastic momentum kicks which occur with a probability which depends on the Doppler shift between the laser and the ion resonance to simulate photon scattering events. By running the simulation a large number of times, the average classical amplitudes of each mode can be found, which can then be converted to mean quantum phonon occupation numbers. Results of such simulations for laser cooling of a six $^9$Be$^+$ ion honeycomb lattice with lattice constant $15$ \textmu m with the confining axes tilted at $ \Theta = 20^\circ$ with respect to the radial plane are shown in figure \ref{fig:coolarray}. Here $B_0$ = $2.5$ T  and $\omega_z = 2\pi \cdot 2.1$ MHz. The uniform laser beam is oriented parallel to the electrode plane so that $\mathbf{k_L} = \cos \Theta \hat{\bf e}_x + \sin \Theta  \hat{\bf e}_z $. The axialization voltage here is $\phi_{\rm ax} = 0.03 \phi_0$. The initial quantum numbers for each mode are chosen within a range close to $10^4$~quanta. The results show that modes can be cooled with time constants in the range of 0.1-0.2 ms, with final mode occupations in the range of 10-30~quanta. This would be expected for Doppler cooling at these trap frequencies. For coupled ion arrays, it is important that the axialization drive mixes all magnetron modes with the modified cyclotron modes. This implies that the modulation strength must be greater than the width of the relevant spread of frequencies of each set of modes.

%\begin{figure}[h]
%  \centering
%  \begin{tabular}{c}
%    \resizebox{7cm}{!}{\includegraphics{freq_hex_6_xy0.3490715.tex}} \\~\\
%    a. Mode frequencies \\~\\~\\
%   \resizebox{7cm}{!}{\includegraphics{cooling_hex_6_xy0.3490715.tex}} \\~\\
%   b. Final average amplitudes and phonon numbers achieved
%\\
% \end{tabular}
% \caption{Normal mode frequencies of the six $^9$Be$^+$ ion honeycomb lattice considered along with results for laser cooling of each mode }  \label{fig:coolarrayfinal}
%\end{figure}

\section{Quantum simulation in fixed lattices} \label{sec:Qsim}
One of the primary uses of multi-ion control in recent years has been for quantum simulation of lattice models. This involves the use of always-on couplings between either spins or motional degrees of freedom of the different ions, realizing a representation of a Hamiltonian of interest elsewhere in physics. Although proposals and experimental demonstrations exist for spin-boson and pure bosonic systems \cite{08Porras, Barreiro2011, Bermudez2011}, a particular focus of recent work has involved Ising spin models implemented on the internal degrees of freedom \cite{12Britton, 17Zhang}. In the following section, we describe how Penning trap arrays could be used for such studies, including optimal electrode layouts, the relevant features of the normal modes, and the implementation of tuneable range Ising spin interactions. These ingredients form a hierarchy; lattice, modes and spin Hamiltonians which are largely shared with other types of quantum simulations which might be of interest. A strength of the systems considered here is that the lattice can be designed through the electrode layout.

\subsection{Optimal Electrode Geometries} \label{sec:electrodes}
From equation \ref{eq:Omex} above, it is clear that the $1/{R^3_{ij,0}}$ nature of effective couplings between the ions favors forming closely-spaced ion arrays. Although this can be achieved by scaling the size of the whole trapping structure, this results in a reduction of the ion-electrode distance which is undesirable due to the expected increase in ion motional heating \cite{06Deslauriers,15Brownnutt}, and the increased chance that stray scattered light from the optical control fields used for cooling and engineered spin-spin interactions induces charging of the electrode surfaces resulting in stray electric fields. For operation of the system, it is also desirable to work with trap frequencies which are high enough to avoid common noise sources in the laboratory, which reduces heating and facilitates laser cooling. For a given electrode structure, the motional frequencies can be increased with a corresponding increase in the applied electrode voltages. However at some point this will be limited by voltage breakdown, and therefore it is beneficial to search for optimal electrode layouts for achieving closely spaced ion traps while retaining high trap frequencies of the individual micro-traps. We consider here the experimental feasibility to generate such surface-electrode trap layouts with high motional coupling between ions in micro-Penning trap arrays for a given applied voltage. Our focus lies in particular on single layer surface-electrode traps as they offer an open planar structure which facilitates optical access. We note that approaches with two planes of electrodes facing each other might allow improved conditions with regards to spin-spin couplings, but these seem to be more technically challenging \cite{14Krauth}.

Previous work has described methods for obtaining surface-electrode geometries which maximize the achievable curvature of the pseudo-potential in arrays of r.f.~traps for a given trapping site density with distance from the electrode surface $h$ \cite{09Schmied}. The problem reduces to maximizing the quadrupole strength which can be produced at the array of sites, which is the identical problem for the Penning trap arrays. However in the case of r.f.~traps this quadrupole potential must be converted into a ponderomotive pseudopotential while maintaining conditions suitable for stable motion, while the Penning trap frequencies are directly dependent on the static quadrupole. The advantage this gives can be evaluated by considering the effect of modulating a static potential $\Pi$ with curvature tensor $\Pi^{(2)} \equiv \partial_\mu \partial_\nu \Pi$, with $\nu, \mu = x,y,z$ at a radio-frequency $\Omega_{\rm RF}$, creating $\Pi \cos(\Omega_{\rm RF} t)$. In the pseudopotential approximation \cite{98Wineland2}, the curvature tensor of the pseudopotential for an ion of mass $m$ is then
\begin{equation}
\Psi_{\rm RF}^{(2)} = \frac{e}{2m \Omega^2_{\text{RF}}} \left[ \Pi^{(2)}  \right]^2
\end{equation}
at any trap center. For simplicity we focus on a cylindrically symmetric trap potential defined as
\begin{equation}
\Pi = \frac{\phi_0}{h^2} \left( z^2 - \frac{x^2 + y^2}{2} \right) \ .
\end{equation}
To compare the strength achievable for the r.f.~vs Penning trap, we take the Frobenius norm of the curvature tensor, finding that the magnitudes of the two curvature tensors can be related using the Mathieu parameter $q_z = -4 e \phi_{0} / \left( m \Omega^2_{\text{RF}} h^2 \right)$ as
\begin{equation}
\lVert \Psi^{(2)}_{\text{RF}}  \rVert
= \frac{\sqrt{3}}{8} \lvert  q_z  \rvert \cdot \lVert  \Pi^{(2)} \rVert .
\end{equation}
Thus the curvature of the pseudopotential is weaker than that of the corresponding static potential by a factor $\sqrt{3} \lvert q_z \rvert /8$. For surface-electrode r.f.~traps, stability becomes a concern for $ \lvert q_z \rvert \gtrsim 0.3$, corresponding to a reduction factor of around 15. In this case, for a given voltage which is applied to the electrodes, the trap frequency in the r.f.~trap is reduced relative to a Penning trap with the same geometry by a factor $\sqrt{15}$.

\begin{figure}
  \centering
  \begin{tabular}{c}
  \LARGE
\resizebox{\columnwidth}{!}{ \includegraphics{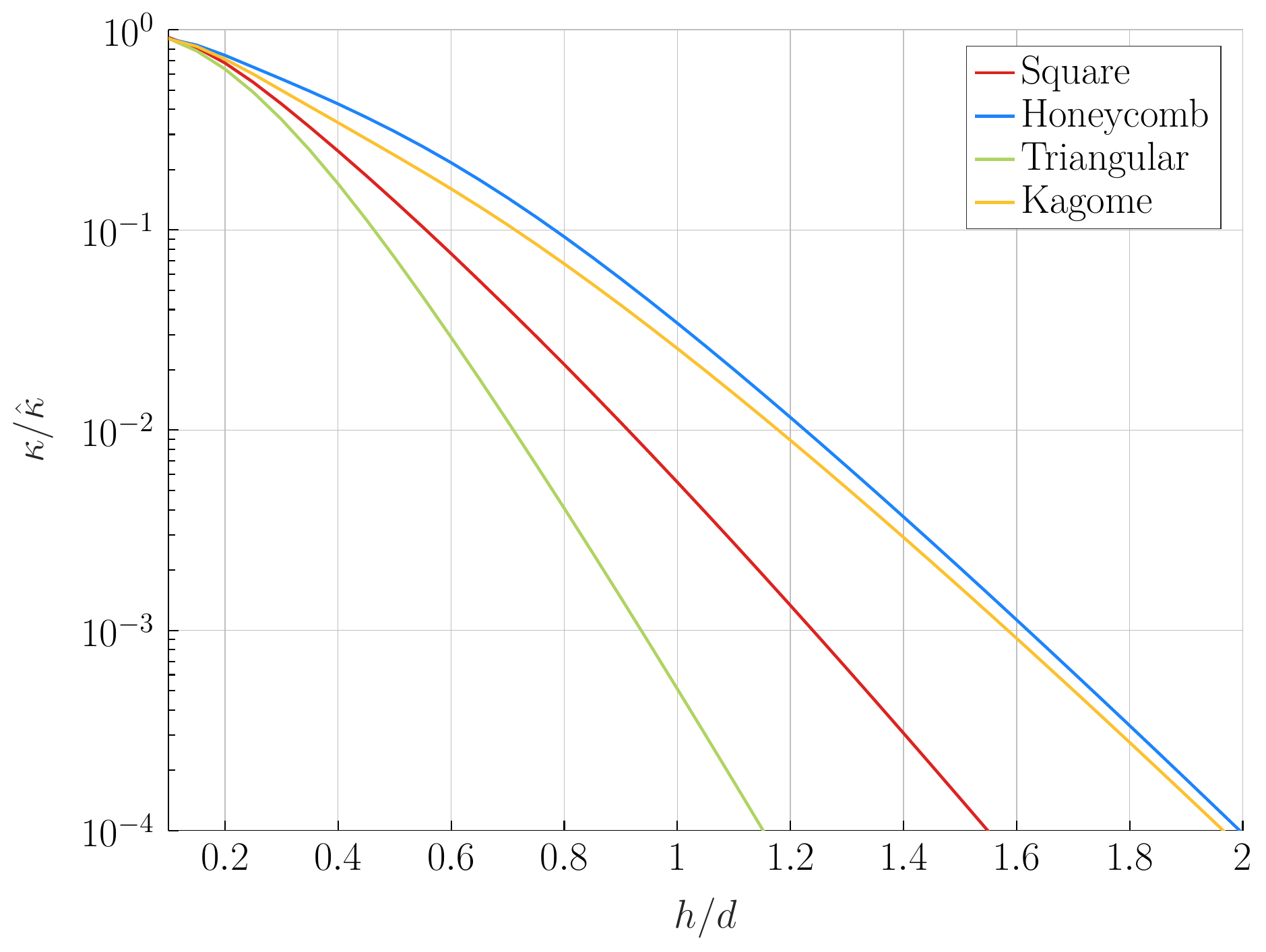}}
 \\

\end{tabular}
\caption{Dimensionless curvatures $\kappa$ as functions of the ratio of microtrap height $h$ to inter-ion spacing $d$, for several lattice geometries.}
\label{fig:curvaturelattices}
\end{figure}

The discussion above makes it clear that the optimization of electrode structures for Penning trap arrays is identical to the case of radio-frequency traps, and thus produces identical electrode geometries \cite{09Schmied,11Schmied}. Similar to the earlier work, we define a dimensionless curvature $\kappa = \lVert  \Phi^{(2)}  \rVert h^2/V$, where $V$ is a fixed voltage applied to part of the electrode plane (the rest being set to zero), and $h$ is taken to be the distance from the centre of the quadrupole to the nearest trap surface. For the symmetric potential the component of the dimensionless curvature aligned with the magnetic field is $\hat{\kappa}_z = \hat{\kappa} 2^{2/3}$. We then optimize the electrode geometry. Figure \ref{fig:curvaturelattices} shows the the dimensionless curvature of the trap potential achievable for different infinite lattices as a function of the ratio of trap height to inter-ion distance, with the magnetic field directed perpendicular to the plane of the electrodes. The values given are normalized to the dimensionless curvature $\hat{\kappa} \approx 0.473$  achievable for an optimal surface-electrode point trap \cite{09Schmied}.

%\begin{figure}
%  \centering
%  \begin{tabular}{c}
%    \LARGE
% \resizebox{\columnwidth}{!}{ \includegraphics{plotkappaanglediffheights.pdf}}
% \\
%
%\end{tabular}
%\caption{Dimensionless curvatures $\kappa$ as functions of the angle of tilt of the trapping axis with respect to the normal of the plane, for several lattice geometries. Here the ratio $h/d$ for each lattice is fixed so that for no tilt we get $\kappa/\hat{\kappa} \approx 10^{-4}$.}
%\label{fig:curvatureangle}
%\end{figure}

\begin{figure}
  \centering
  \begin{tabular}{c}
    \LARGE
 \resizebox{\columnwidth}{!}{ \includegraphics{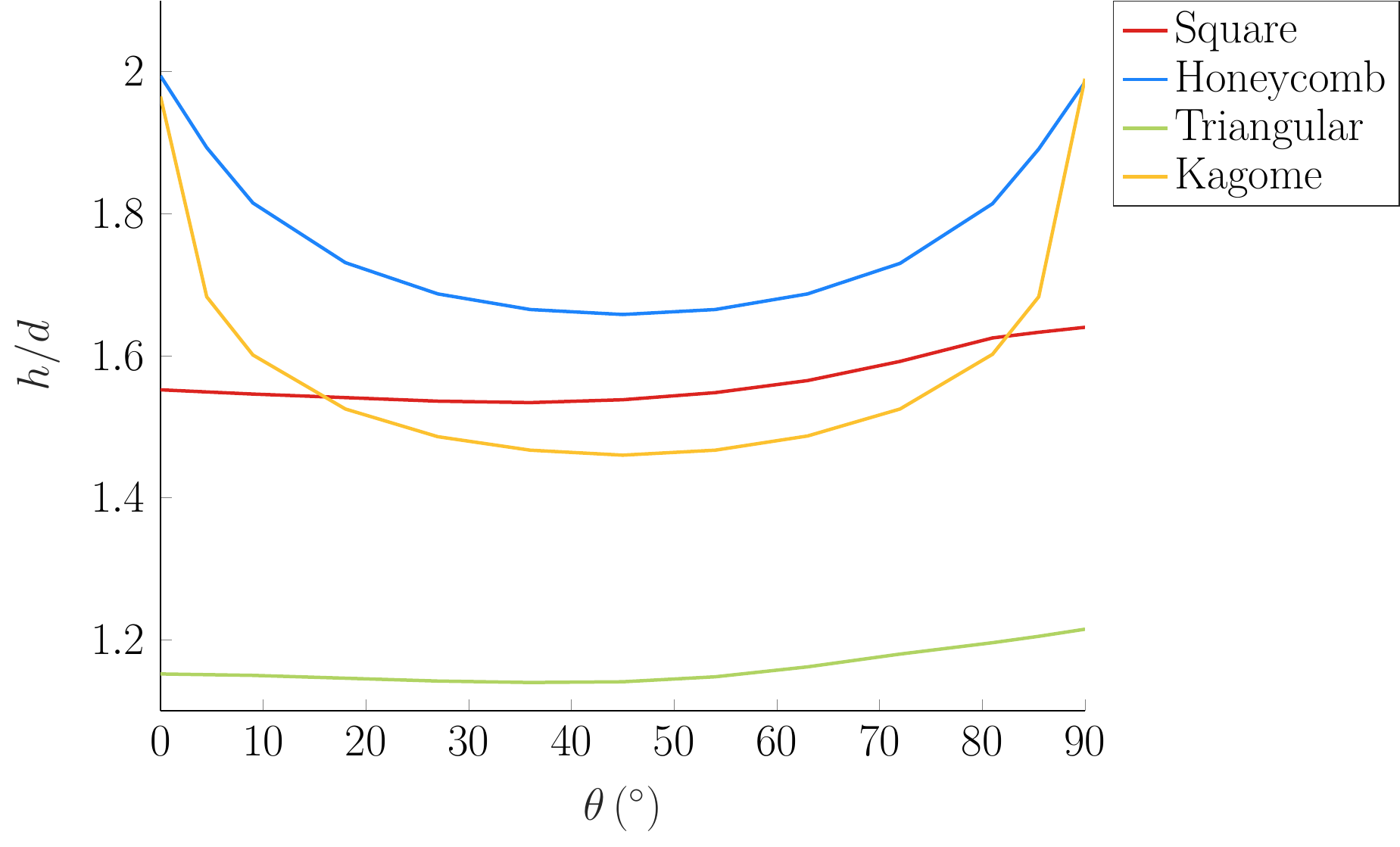}}
 \\

\end{tabular}
\caption{The value of $h/d$ for which a dimensionless curvature $\kappa/\hat{\kappa} = 10^{-4}$ is achieved as a function of the angle of tilt of the trapping axis with respect to the normal of the plane, for several lattice geometries. For the hexagonal and Kagome lattices a strong effect is observed.}
\label{fig:curvatureangle}
\end{figure}

Typical electrode structures in quantum information experiments can withstand differences in voltages on neighboring electrodes sufficient to achieve $V = 300$~V \cite{13Sterling}. With $h = 30$~\textmu m, $\kappa = 10^{-4}$ then allows to achieve $\omega_z = 2 \pi \cdot 2.1$~MHz for beryllium ions. Figure \ref{fig:curvaturelattices} indicates that this would allow ions spaced by between $26$~\textmu m and  15~\textmu m, with the former corresponding to the triangular lattice and the latter to the  honeycomb and Kagome lattices. The resulting exchange frequencies are between 11~kHz $< \Omega_{\rm ex, z}/(2 \pi) < 55$~kHz. This is far above heating rates and frequency drift rates observed in traps of a similar size, thus high quality coherent exchange would be expected \cite{11Brown}.

The discussion above considers trapping potentials with the confinement direction (and magnetic field) out of the plane of the electrodes. The introduction of a tilt between the electrode plane normal and the magnetic field modifies the geometries of the optimized electrodes and the values of $\kappa$ which can be achieved. Figure \ref{fig:curvatureangle} shows the value of $h/d$ for which a dimensionless curvature $\kappa/\hat{\kappa} = 10^{-4}$ is achieved as a function of the angle $\Theta$ between the magnetic field and the normal to the electrode plane. We see that $\Theta = 90^\circ$ produces the highest curvatures for all geometries (and thus requires the smallest $h/d$ to get to $\kappa/\hat{\kappa} = 10^{-4}$), with an additional maximum for $\Theta = 0^\circ$. The former allows laser cooling with laser beams in the plane of the electrode surface, the latter does not - previous work in radio-frequency traps indicates that $\Theta > 8^\circ$ is necessary for robust laser cooling \cite{17Lindenfelser}. The change in achievable $h/d$ for a fixed $\kappa$ between $\Theta = 0^\circ$ and $\Theta = 8^\circ$ is small for the triangular and square lattices, but considerable in the case of the Kagome and honeycomb lattices.

\subsection{Spin-spin interactions for the Ising Model} \label{sec:spinspin}
We now examine the possibility of implementing spin-spin couplings of the form relevant to studying models such as the transverse Ising model \cite{25Ising}, which has been a common target of quantum simulation experiments using trapped ions. An effective spin-spin interaction can be generated based on standard techniques developed in the trapped-ion quantum computing community to implement multi-qubit gates \cite{03Leibfried,99Molmer,08Roos,09Kim}. These methods rely on the application of forces that depend on the internal (pseudo)spin-state of the ions. For instance, two laser beams off-resonant with respect to the internal transition and with a frequency difference $\mu_R$ and wave-vector difference $\mathbf{k}_R$ between each other create a traveling-wave interference pattern at the ions. Each ion experiences a state-dependent optical dipole force (ODF) that oscillates at the frequency $\mu_R$. To simplify the algebra, we assume that the two relevant states of the ions are spin-half ground states with no hyperfine structure \cite{06Home}. In this case, it is possible to generate an ODF that is equal in magnitude but opposite in sign for the two internal states which can be considered as eigenstates of the Pauli operator, $\hat{\sigma}^z$. For small coherent displacements of the ions from their equilibrium positions we can use the Lamb-Dicke approximation and keep only resonant terms, resulting in the ODF interaction Hamiltonian
\begin{equation}
\hat{H}_{\text{ODF}} \approx  \sum^N_{j=1} E_O \mathbf{k}_R\cdot\mathbf{r}_{j} \sin(\phi_j - \mu_R t)\hat{\sigma}^z_j ,
\label{eq:ODF}
\end{equation}
where $E_O$ depends on the laser beam properties as well as the matrix elements of the internal transition of the ions, and the phase at the ion location is given by $\phi_j = \mathbf{k}_R\cdot\mathbf{R}_{j0}$. A similar Hamiltonian can be achieved in a rotated spin basis ($\hat{\sigma}^x, \hat{\sigma}^y$) by driving both red and blue motional sidebands of the spin-flip transition simultaneously \cite{08Roos}. Given the periodic arrangement of ions, we can ensure that this phase is the same for all ions using well-chosen laser beam configurations. To simplify the following discussion, we assume that this condition is met (see Appendix \ref{app:SpinSpin} for a more general treatment).

The time evolution operator associated with $\hat{H}_{\text{ODF}}$ can be calculated by carrying out a Magnus expansion in the interaction picture, and for the given ODF interaction this yields two terms. The first term describes periodic spin-motion entanglement generated by the ODF. Quantum simulation experiments typically work in the regime in which this entanglement is negligible and can therefore be adiabatically eliminated \cite{12Blatt}. The second term describes an effective Ising-like spin Hamiltonian
\begin{equation}
\hat{H} _ {\text{SPIN}} = \sum _ { j j ^ { \prime } } J _ { j j ^ { \prime } } ( t ) \hat{\sigma} _ { j } ^ { z } \hat{\sigma} _ { j ^ { \prime } } ^ { z } \ ,
\end{equation}
with the static part of the spin-spin interactions $J _ { j j ^ { \prime } } ( t )$  given by
\begin{equation}
\begin{aligned}
J^0 _ { j j ^ { \prime } }
&=\frac{E^2_O}{2 \hbar} \sum_{\lambda} \frac{\omega_{\lambda}}{\mu^2_R -\omega^2_{\lambda}} \text{Re} \left( \eta^{*}_{\lambda j} \eta_{\lambda j^{\prime}} \right) \ .
\end{aligned}
\end{equation}
Here we have defined the Lamb-Dicke parameters in a slightly unconventional manner as $\eta_{\lambda j} = \sum_{\nu = x,y,z} k^{\nu}_{R} \rho_{\lambda 0} \gamma_{\lambda j \nu}$, with $\rho_{\lambda 0}$ being the spread of the zero-point wavefunction of mode $\lambda$ and $\gamma$ is the corresponding eigenvector normalized to 1. For an r.f.~trap this definition reduces to the standard form of the Lamb-Dicke parameter \cite{11Home}.

 \begin{figure}
  \centering
    \begin{tabular}{c}
    \LARGE
  \resizebox{7cm}{!}{\includegraphics{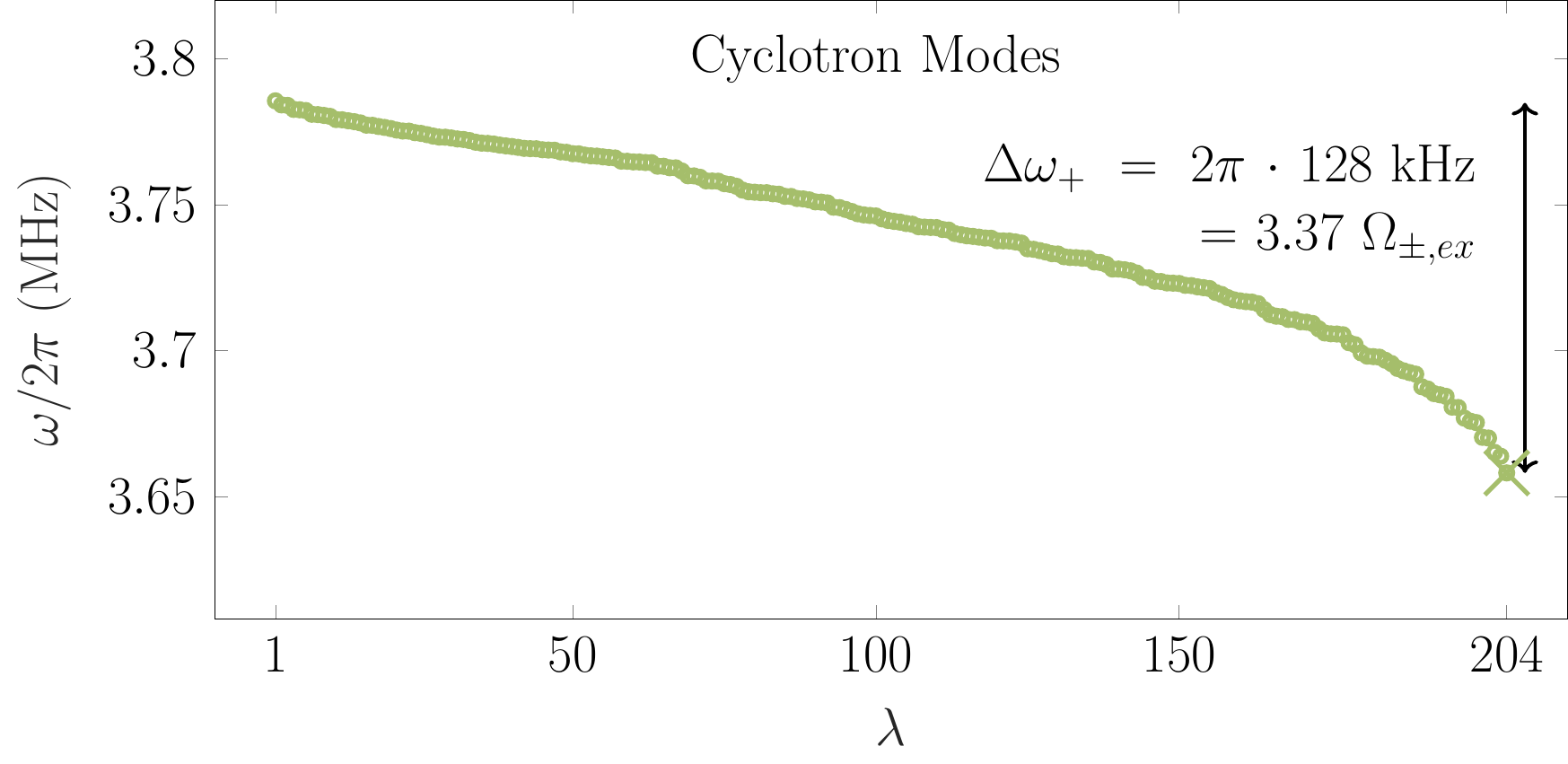}} \\
   \LARGE
  \resizebox{7cm}{!}{\includegraphics{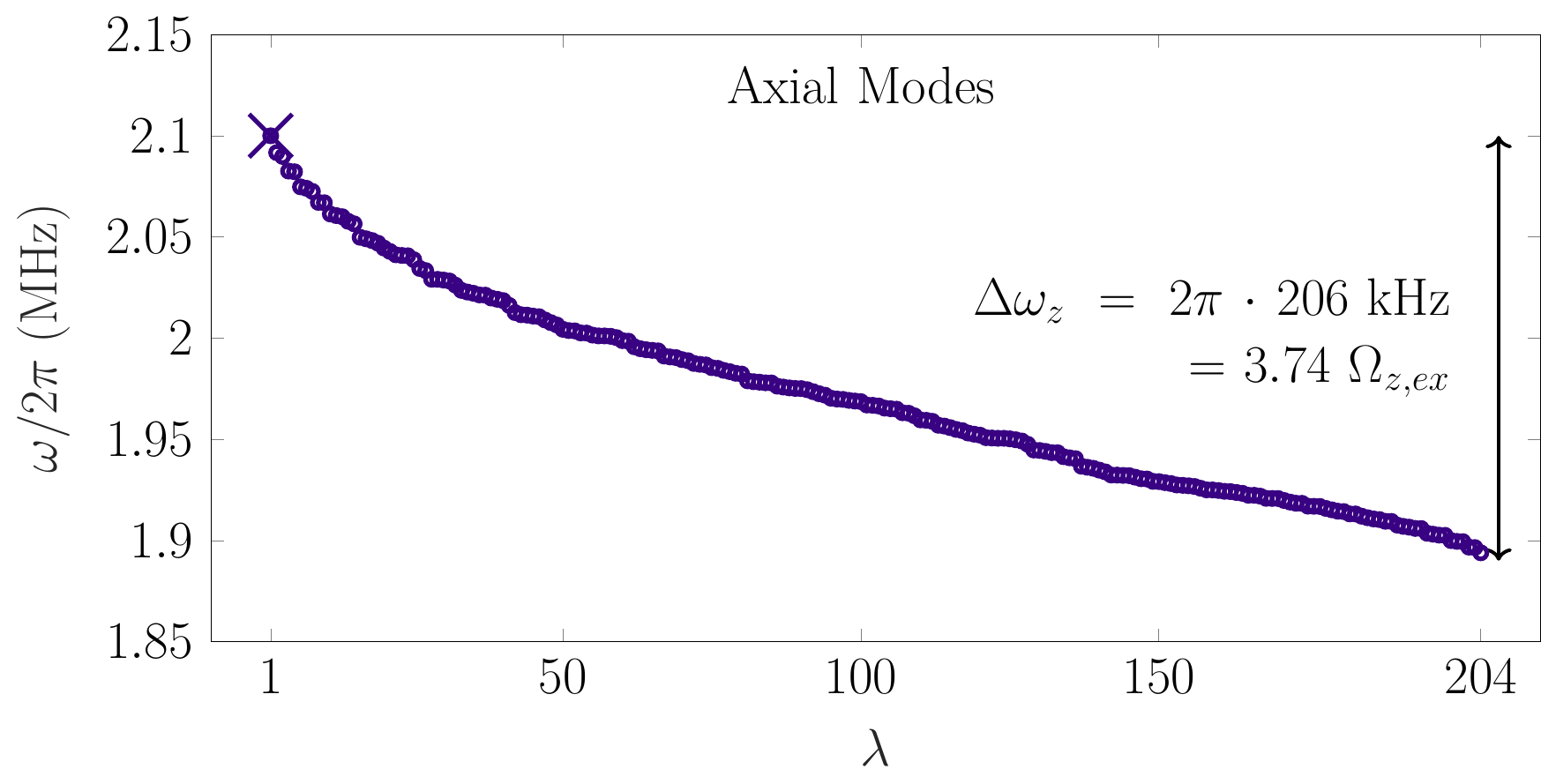}} \\
   \LARGE
 \resizebox{7cm}{!}{\includegraphics{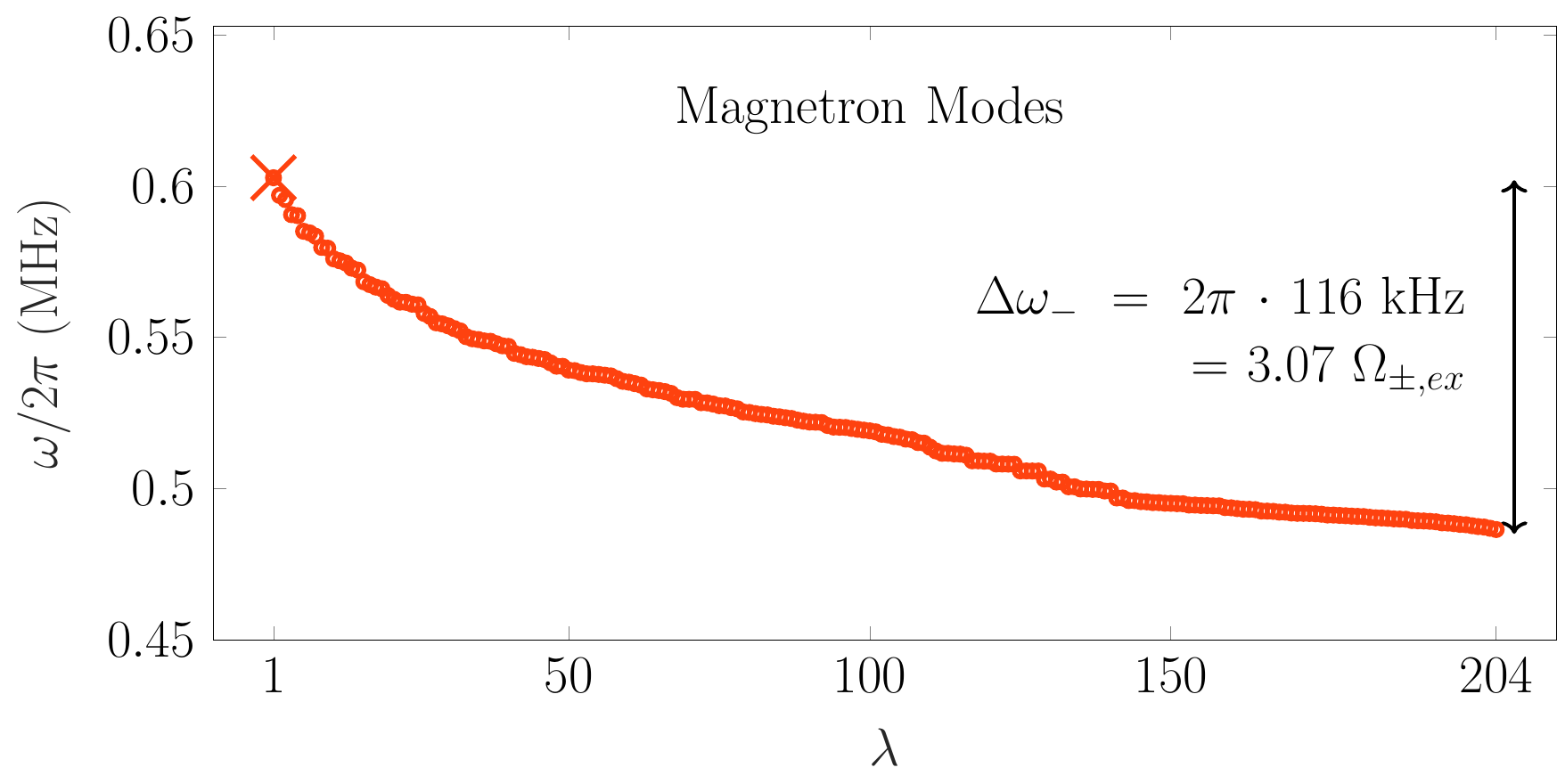}}
 \end{tabular}
 \caption{Frequency spectrum of a 204 ion honeycomb lattice (nearest-neighbor distance $d = 15$ \textmu m) arranged with the tilted configuration ($\Theta = 20^\circ$). Shown here are the cyclotron, axial and magnetron branches of the normal modes. The centre of mass frequency in each mode branch is marked with a cross. The width of each branch is shown in terms of the respective 2-ion exchange frequency through black arrows.}\label{fig:modesB}
\end{figure}

An interesting aspect of the simulation of the Ising model using trapped ions is the possibility to engineer spin-coupling terms which follow a power-law scaling with the inter-ion separation $ \lvert J_{j j^{\prime}} \rvert  \propto 1/\lvert R_{j j^{\prime} 0} \rvert^a$, with $a$ dictated by the experimental detunings \cite{12Britton}. However, such tunable-range interactions are possible only with certain mode structure, in which the center-of-mass (COM) mode has the highest (lowest) frequency in a given band, and the state-dependent force is tuned to a higher (lower) frequency than this mode. In previous experiments with bulk crystals this situation is naturally satisfied, whereas it is not always satisfied in the trap arrays which we consider in this paper. In the following, we trace the importance of the normal mode structure in the determination of the effective spin-spin interactions that can be engineered for a given system of ions. We take as an example a honeycomb lattice of 204 ions, with the nearest-neighbor separation  $d = 15$ \textmu m. Here $B_0$ = $2.5$ T  and $\omega_z = 2\pi \cdot 2.1$ MHz. We use an ODF interaction strength corresponding to a Rabi frequency of $E_O / \hbar = 2 \pi \cdot 300$~kHz, which is a level similar to that used in previous experiments \cite{17Zhang}.

%It is evident from this expression that the structure of the normal modes of the trapped ion system is essential to the discussion of the effective spin-spin interactions that can be engineered using a given configuration of ODF-generating beams.
%The micro-Penning trap arrays proposed here form a flexible platform to realize a variety of two-dimensional lattices, and in turn, spin-spin Hamiltonians. It is sufficient, however, to illustrate the general ideas with the help of a two-dimensional triangular lattice with different possible geometric orientations with respect to the common trapping axis. In what follows the magnetic field is always aligned along this axis and for a quadrupolar electric potential that is symmetric along the other two axes it is sufficient to specify the orientation of the quadrupoles through the direction of the magnetic field.

We first consider the case when the magnetic field is aligned normal to the plane in which the ions lie. With such an orientation the axial motion is decoupled from the radial motion and the COM mode sits at the highest frequency in the axial band. With a wave-vector difference $\mathbf{k}_R = k_R \hat{\bf e}_z$, only the axial modes are excited. By tuning the ODF to the blue of the axial branch variable range spin-spin couplings can be engineered with the range of interaction decreasing from infinite range ($a=0$) to dipole-dipole type ($a=3$) as $(\mu_R - \omega_z)$ is increased. Since all coupling terms are positive, this allows to simulate an antiferromagnetic Ising Hamiltonian. Experiments carried so far using both r.f.~traps (for eg. Ref.~\cite{17Zhang}) and Penning traps (for eg. Ref.~\cite{12Britton}) are based on this simplification. Conversely, a tunable-range ferromagnetic Ising model can be simulated by aligning $\mathbf{k}_R$ along the radial plane and tuning the ODF to the red (blue) of the cyclotron (magnetron) branches. This is possible since the cyclotron (magnetron) COM mode is the lowest (highest) in its branch. The coupling terms for the magnetron branch are negative since the magnetron motion represents an inverted harmonic oscillator. %Such a model can not be implemented over different ranges of interaction using 1D systems such as linear r.f.~traps and 2D systems such as ion crystals in Penning traps because of the lack of necessary mode structure with COM modes at the extrema in all phonon branches and the bulk rotation of the crystal in the radial plane respectively.

\begin{figure}
  \centering
  \begin{tabular}{c}
   \large
  \resizebox{\columnwidth}{!}{  \includegraphics{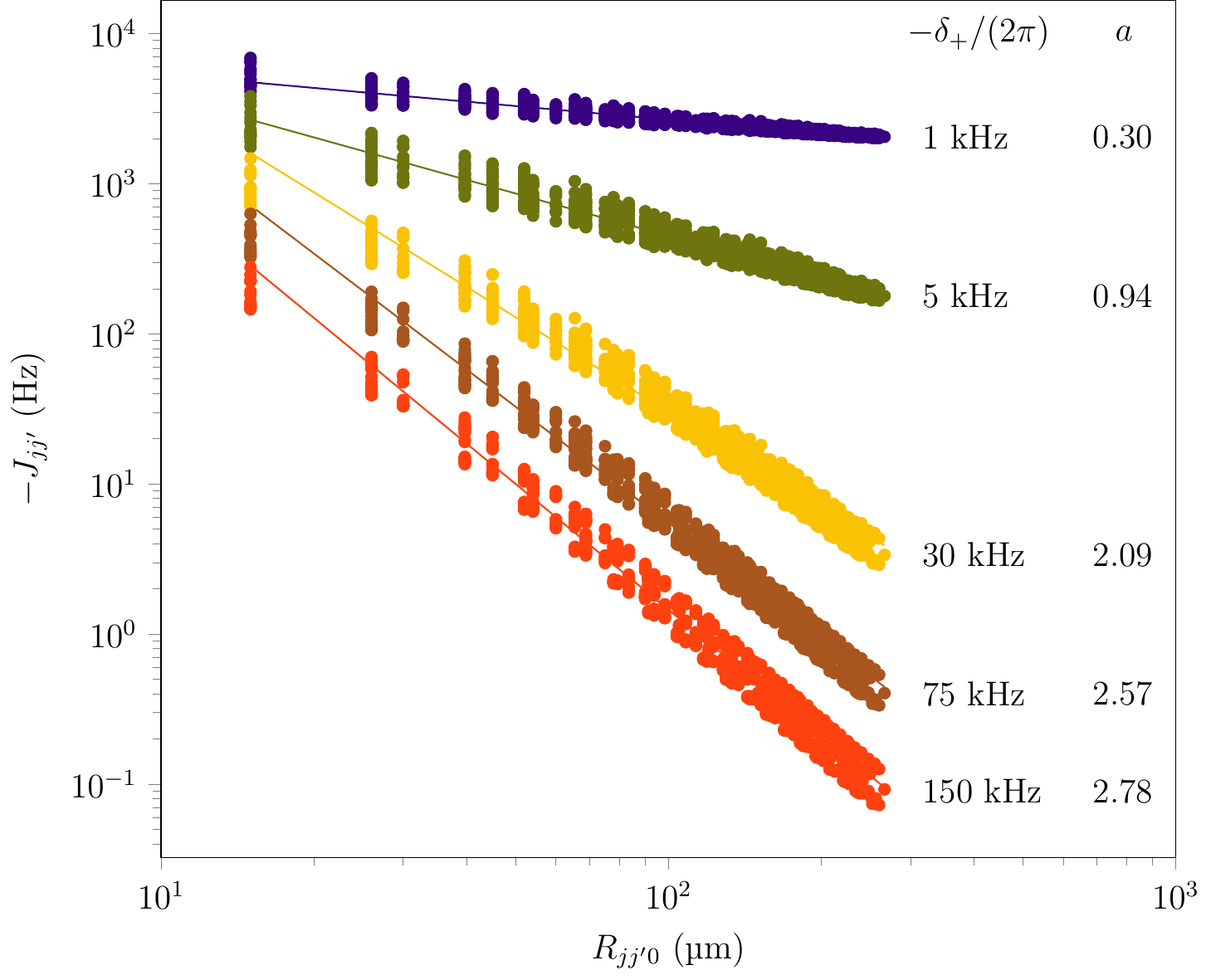} }
 \\
i. Cyclotron Branch
\\~\\
 \large
  \resizebox{\columnwidth}{!}{  \includegraphics{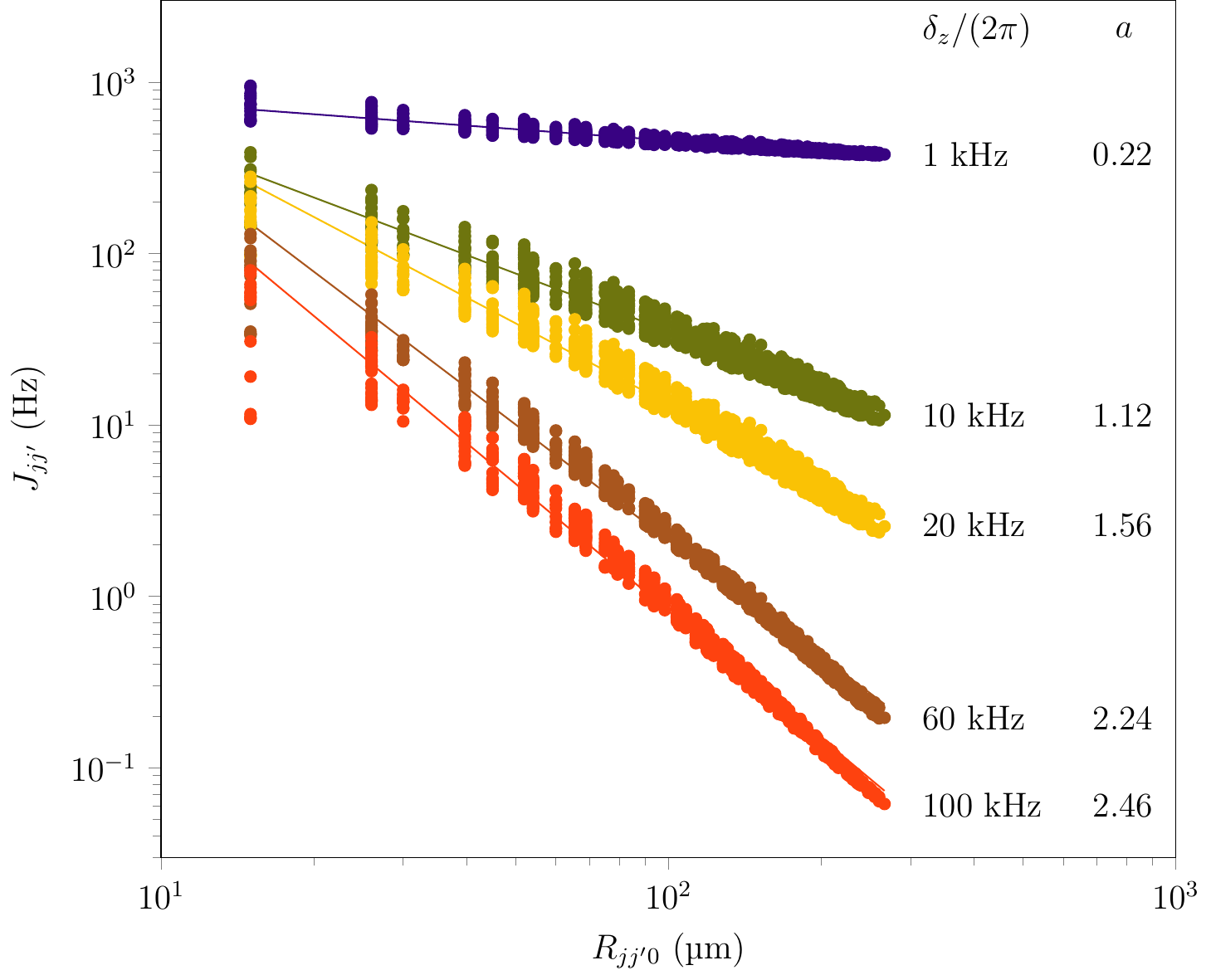} }
 \\
ii. Axial Branch
\end{tabular}
 \caption{Spin-spin coupling terms generated with an optical dipole force for a 204 ion honeycomb lattice ($d = 15$~\textmu m) arranged with the tilted geometry ($ \Theta = 20^\circ$). The ODF can be created by two laser beams along the plane of the electrodes so that the difference wave vector is given by $\mathbf{k}_R = k_R \cos \Theta  \hat{\bf e}_x + k_R \sin \Theta  \hat{\bf e}_z$. When the beatnote frequency $\mu_R$ lies to the red of the cyclotron branch, all couplings are negative and a ferromagnetic Ising interaction can be engineered. These couplings follow an approximate power law decay $\lvert J_{j j^{\prime}} \rvert \propto 1/\lvert R_{j j^{\prime} 0} \rvert^a$ and the exponent $a$ increases with increasing magnitude of $\delta_+ = \mu_R - \omega_+$. Here $E_O / \hbar = 2 \pi \cdot 300$ kHz. Similar tunable ferromagnetic couplings can be achieved by tuning to the blue of the magnetron COM mode. For the radial modes, lasers parallel to the electrode plane can also be used to create a wavevector difference along the $\hat{y}$-axis. When $\mu_R$ is increasingly tuned away from the axial branch so that $\delta_z = \mu_R - \omega_z$ increases, all couplings are positive and hence variable range antiferromagnetic Ising like interactions can be generated with $a$ going from 0 to 3. The strength of these couplings is limited by the angle of tilt $\Theta$. Use of laserless methods could eliminate this restriction} \label{fig:spinspin}
\end{figure}

\begin{figure}
  \centering
  \begin{tabular}{c}
   \LARGE
  \resizebox{\columnwidth}{!}{\includegraphics{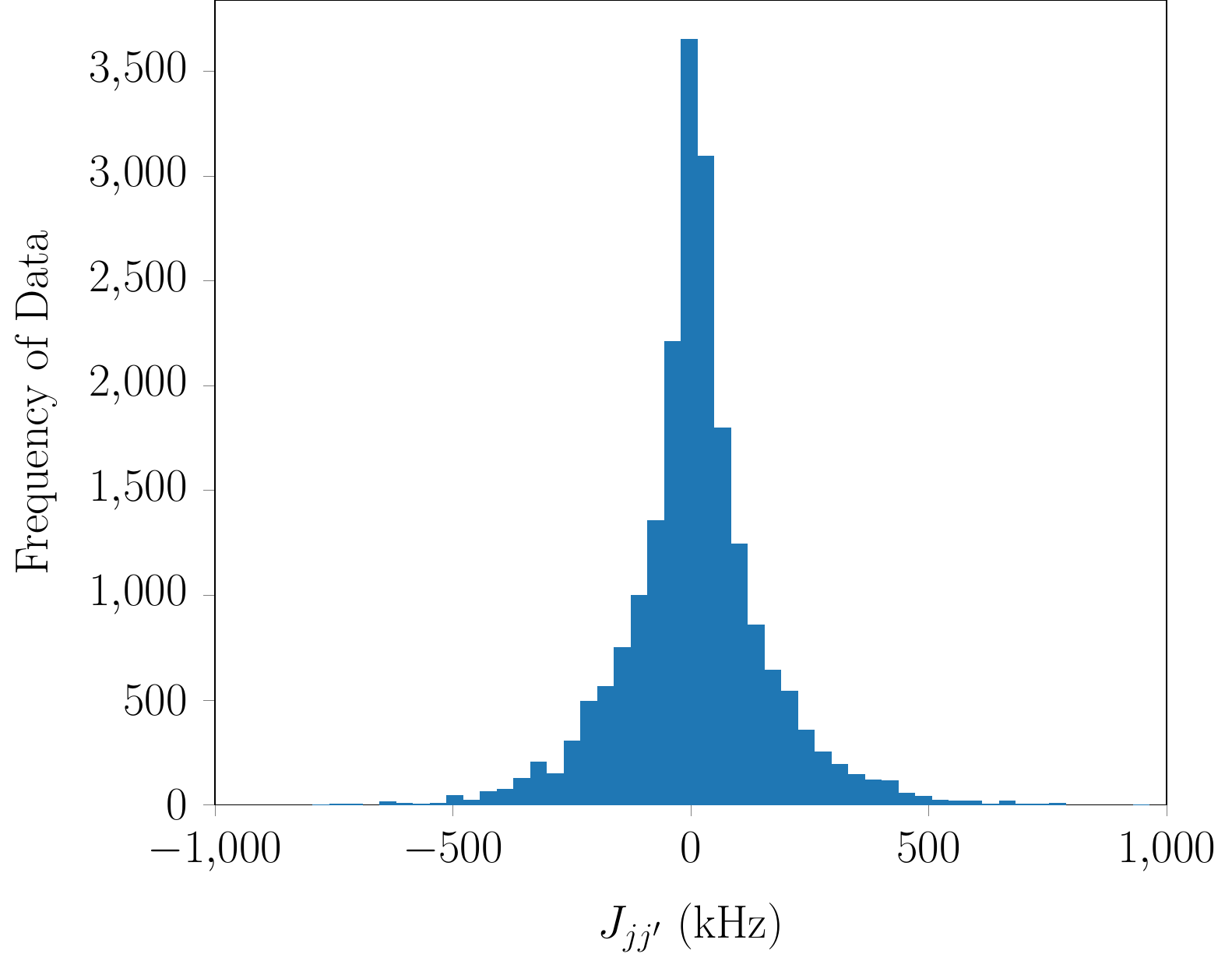}}
   \end{tabular}
 \caption{Histogram of spin-spin coupling terms when the beat note frequency $\mu_R$ is tuned slightly away from the COM frequency of the axial branch. Here $\mu_R - \omega_{z} = 2\pi \cdot 0.5$ kHz}  \label{fig:spinspinhistogram}
\end{figure}

One of the challenges for realizing a setup with the magnetic field normal to the plane is that it is difficult to cool the axial motion or generate an ODF along the axis using lasers directed parallel to the surface of the chip, or equivalently, the plane where the ions sit. While these problems can be countered by using in-chip waveguides \cite{16Mehta}, we rather attempt to see how well the behavior described above holds when the magnetic field is tilted at an angle $\Theta$ with respect to the normal of the lattice plane. The crucial factor here is the position of the COM mode within the branch being excited by the ODF. For the trap settings and lattice considered here, the COM mode is on the edge of the band for $\Theta \lesssim 43^\circ$. Figure \ref{fig:modesB} shows the normal mode spectrum for the case of $\Theta = 20^\circ$, which does satisfy this condition. Spin-spin coupling terms generated from this lattice through tuning the ODF outside the frequency spectrum of the cyclotron and axial branches are plotted in figure \ref{fig:spinspin}.

%It is worth noting that due to the time dependence of the radial components of the equilibrium positions spin-spin interactions using radial modes as has been described in the above two cases can fundamentally not be engineered in quantum simulators based on ion Coulomb crystals. Similarly, for a linear r.f.~trap, only the motion along the axis of equilibration of the ions can be used to implement a variable range Ising model. \footnote{This last sentence is certainly wrong. Both Chris Monroe and Christian Roos use radial modes for their variable range Ising simulations. The axial modes for a long chain would be too low in frequency, and I think also have the wrong style of splitting. For the bulk-crystal Penning trap, I would make the same argument, that the in-plane modes have to be low frequency in order to produce the extended crystal. We do not require this mode-softening, which is a big advantage of our setup. }

For the case of $\Theta = 90^\circ$, that is, when the magnetic field is along the plane, a suitably oriented in-plane laser beam can cool all motional modes. However all three COM modes lie away from the extrema of their respective branches, making it hard to implement variable range spin-spin interactions by tuning $\mu_R$ alone. With such a mode structure, detuning $\mu_R$ in either direction from any of the three COM modes does not reveal a well-defined power law decay and the coupling terms have different signs depending on the angle of the inter-ion vector. A histogram plot in figure \ref{fig:spinspinhistogram} shows this behavior. The same is expected more generally whenever $\mu_R$ lies in the middle of the branch of modes used. More complicated methods involving multi-frequency laser beams could make it possible to emulate a tunable-range Ising model with such geometric arrangement of micro-traps \cite{12Korenblit}, but we do not consider this here. At large detunings of the ODF from a given branch, dipole-dipole couplings can be realized with a distance scaling $ \lvert J_{j j^{\prime}} \rvert  \propto 1/\lvert R_{j j^{\prime} 0} \rvert^3$ but the sign of the coupling term between any given pair of spins will be determined mostly by their relative phase in the mode closest to the chosen value of $\mu_R$. This frustration in sign might allow for the study of disordered spin dynamics in poorly understood systems such as quantum spin glasses \cite{75Edwards}. The same behavior could also be effected by tuning within any phonon branch although the emergence of any power law scaling is not expected, except for the case when the ODF is tuned close to the COM mode, leading to infinite range behavior ($a=0$).

While the statements above give a general discussion, they do not include the tuning of the angle between the projection of the B-field into the plane and the lattice symmetry axes. Here it is probable that special cases arise with interesting features. This is an area of future study.

\section{Quantum computation with local modes} \label{sec:Qcomp}
Many of the most promising approaches to quantum error-correction also make use of extended two-dimensional lattices of qubits.
These include both the surface code \cite{Fowler2012} and the topological color codes \cite{Bombin2006}.
For fault-tolerance errors must be local and must have low rates, thus it is desirable that gates between ions involve only the chosen ions and do not require precise control of the complete array.
For this reason it is desirable to decouple these chosen ions from the rest of the array.
In our architecture this can be achieved by local tuning of the trap potential of the ions in question.

\begin{figure}
  \centering
  \begin{tabular}{c}
   \LARGE
  \resizebox{0.7\columnwidth}{!}{\includegraphics{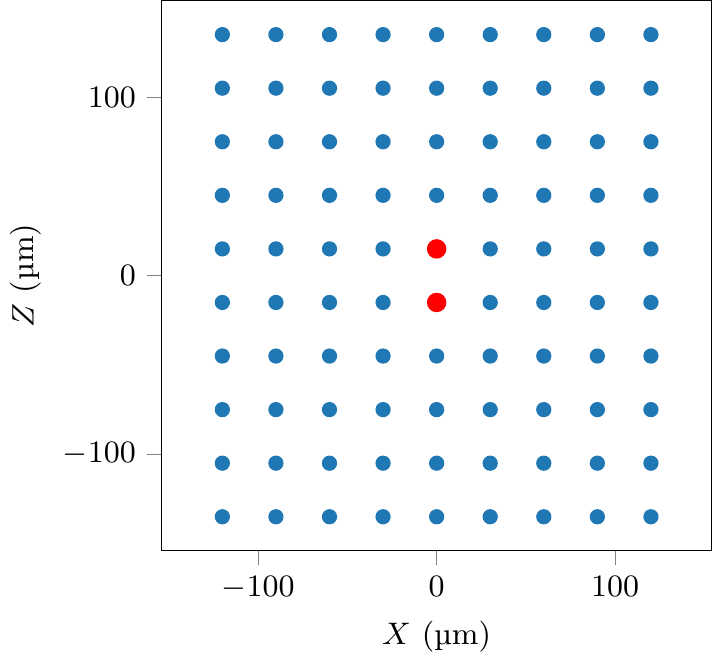}} \\
  i. Lattice\\~\\
   \LARGE
  \resizebox{0.9\columnwidth}{!}{\includegraphics{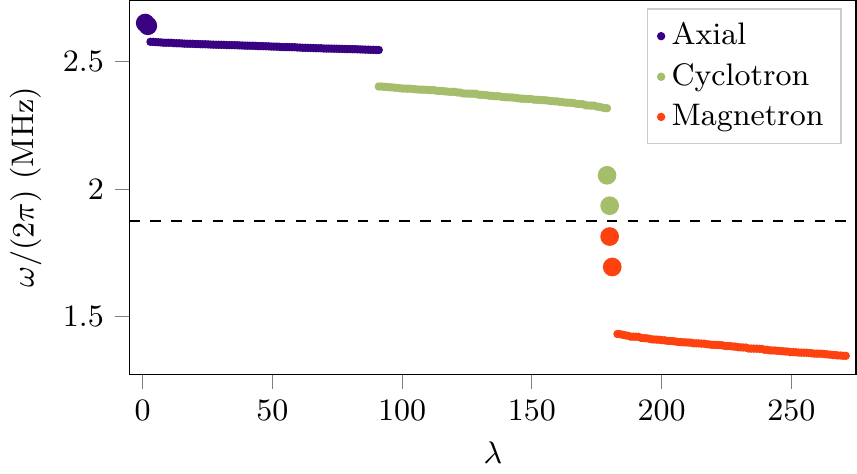}} \\
  ii. Normal mode spectrum
   \end{tabular}
 \caption{
i. 90 ion square lattice of beryllium ions, with two selected ions (shown larger) in a potential well with a different curvature from all of the others, achieved by local tuning of the electrode voltages.
ii. Normal mode frequencies for the lattice considered.
Modes 1-90 are axial modes, 91-180 are modified cyclotron modes and 181-270 are magnetron modes.
The isolated modes of the ions at two selected sites are shown with enlarged symbols, corresponding to mode indices 1 and 2, 179 and 180  and 181 and 182.
Due to their frequency separation from the bulk these modes are largely isolated.
The horizontal dashed line indicates half the bare cyclotron frequency $\omega_c/2$.
} \label{fig:compmodesplot}

\end{figure}
As a test case, we consider a lattice of 90 beryllium ions on a square lattice spaced by 30 \micron~with a quadrupole potential corresponding to $\omega_z = 2 \pi \cdot 2.55$~MHz (see figure \ref{fig:compmodesplot} i).
We choose a magnetic field of 2.2~T which lies in plane, for which
$\omega_c/(2 \pi) = 3.75$~MHz.
For an ion in a single isolated potential, the modified-cyclotron and magnetron modes are then at $2.39$~MHz and $1.36$~MHz respectively.
In order to perform a gate locally between the central two ions, we tune the curvature of the potential for these two ions such that the axial oscillation frequency for a single isolated ion is $(\omega_z + \Delta \omega_z) = 2\pi \cdot 2.63$~MHz, where $\Delta\omega_z = 2 \pi \cdot 80$~kHz.
The resulting mode spectrum is shown in figure \ref{fig:compmodesplot}.
The choice of curvature is special, because the potential it produces meets conditions for which the resulting modes have a relatively small separation between the uncoupled magnetron and modified-cyclotron modes (a contribution to which is made by the potential of the neighboring ions).
Since the zero-point motion of these modes scales as $1/ \Delta_{\alpha} $ with $\Delta_{\alpha} = \omega_\alpha - \omega_c/2$, such a choice enhances the zero-point motion of the modified cyclotron and magnetron modes.
In addition, these modes are relatively far-detuned from the modes of the rest of the lattice.
For the chosen curvature, the coupled modes closest to $\omega_c/2$ are those in which the selected ions move out of phase (often called stretch modes), which have frequencies for which $\Delta_{\rm s,+} = 2\pi \cdot 60.2$~kHz and $\Delta_{\rm s,-} =  2 \pi \cdot - 60.50 $~kHz  and zero-point motion for the chosen ions of $96.4$~nm.
The next closest are the center-of-mass (COM) modes of the selected ions, with $\Delta_{\rm c, +} = 2\pi \cdot 179.3$~kHz and $\Delta_{\rm c,-} = 2 \pi \cdot -179.6$~kHz with corresponding zero-point motion of $55.9$~nm.
These then form an isolated set of modes on which a multi-qubit gate can be performed.
 Similar to the approach taken in section \ref{sec:Qsim}, we consider a geometric phase gate \cite{03Leibfried, 06Home} which uses an oscillating state-dependent force such as can be produced with a traveling standing wave or  a magnetic field gradient, and make the simplification that the force at any given point is equal in magnitude and opposite in sign for the two eigenstates of $\hat{\sigma}_z$.
The Hamiltonian is then the one found in equation \ref{eq:ODF}, and we assume that the experiment can be arranged such that the phase of the force is the same at each ion.
To perform a gate, the frequency $\mu_R$ must be chosen.
For the mode detunings above, an attractive possibility is to use an oscillating force at $\mu_R = \omega_c/2$, which drives the two stretch modes almost equally, but with the opposite detuning.
While for a Paul trap this would result in the phases due to each mode cancelling out, for the Penning trap the contributions of both modes add, because the lower frequency mode of the pair is a magnetron mode and thus has a negative frequency.
If only these two modes were included in the gate, it could be performed in a time of $t_g  = 2 \pi / \Delta_{s,+}$ by using a Rabi frequency $\eta_s \Omega = \Delta_{\rm s, +}/\sqrt{2}$.
In practice the contributions of the COM modes subtract from this effect, and the additional bulk modes also contribute.
For a Lamb-Dicke parameter of $\eta\sim 0.17$ for the stretch modes, simulations involving all modes show that a Rabi frequency of $E_{O} / \hbar = 2 \pi \cdot 300$~kHz  could be used to perform a gate which would produce a Bell-state with fidelity of $F > 0.9998$ in 16~\textmu s.
The large zero-point motion means that a Raman beam pair with a small difference wavevector is required to operate within the Lamb-Dicke regime, which is desirable for insensitivity of gate fidelity to initial ion temperature \cite{03Leibfried}.
For beryllium this would require an angle of $ \theta_R = \pi/36$ for the beryllium wavelength of $313$~nm.
Alternatively a magnetic field gradient of $\sim 19$~T/m (lower than that realized in recent experiments \cite{Khromova2012}) would provide a gate with the same speed.
Note that the motional mode parameters used in this analysis were chosen to satisfy a close-to-integer ratio between $\Delta_{\rm c}$ and $\Delta_{\rm s}$, such that both the local stretch and COM modes are dis-entangled from the internal states at the end of the gate \cite{03Leibfried}.

These results indicate that local changes to the potential, combined with individual optical or microwave addressing of the ions could be used to realize quantum computing in the proposed architecture.
The enhanced zero-point motion used here is particularly appealing in the context of magnetic field gradient gates, which struggle to achieve high gate speeds in Paul trap settings due to the challenge of producing high field gradients \cite{11Ospelkaus, Harty2016, Weidt2016, Khromova2012}.
In the presence of high Rabi frequencies, faster gate speeds should be possible using multi-pulse techniques following methods demonstrated in Paul traps \cite{Steane2014, Schaefer2018}.
For error-correction the need for regular detection of ancilla ions poses challenges with regards to measurement cross-talk, which might require the use of selective electron shelving \cite{98Wineland2}.
For parallelizing error-correction codes, it would be necessary to select multiple pairs of ions at different points in the lattice, and perform gates on each of these simultaneously.
Here the $1/d^3$ nature of the Coulomb mode coupling is advantageous.

\section{Implementation example}
\label{sec:example-implementation}

\begin{figure*}
\centering
\LARGE
\resizebox{2\columnwidth}{!}{\includegraphics{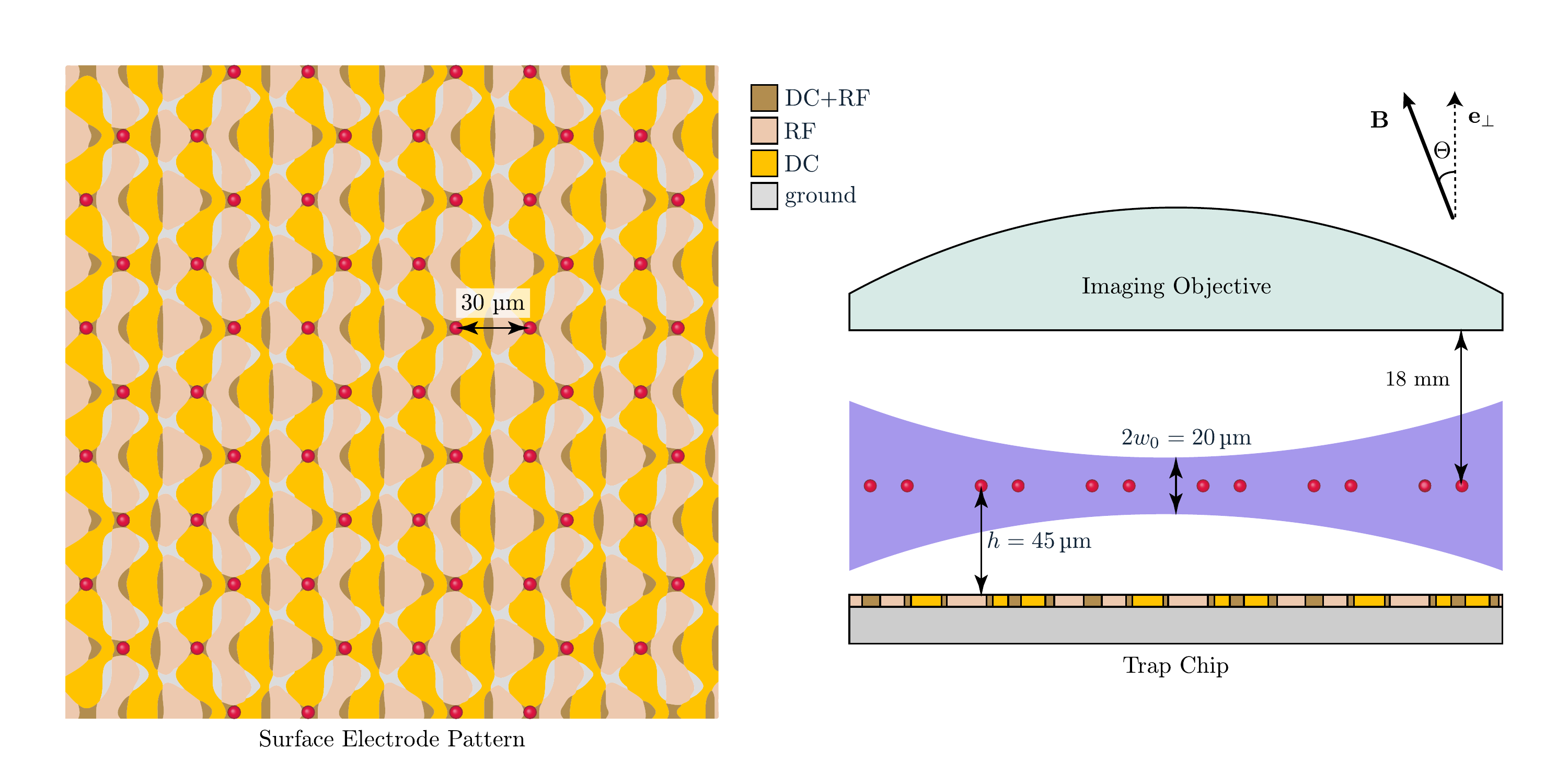}}
\caption{
  Top and side view of an example trap architecture for a honeycomb array of 62 beryllium ions, with a nearest-neighbor distance $d = 30$ \micron.
  Ions are trapped at a distance of $h = 45$ \micron~above the surface of the electrodes and are globally addressed, cooled, and detected by lasers parallel to the surface.
  The electrodes labeled DC generate the static electric potential, which in conjunction with the global magnetic field, provides 3-dimensional confinement at the sites marked in red.
  Tilting the magnetic field and the confining axis of the electric potential with respect to the normal of the electrode plane allows for laser cooling of all modes of motion with lasers parallel to the surface.
  The electrodes labeled RF provide the weak oscillating axialization field for laser cooling the radial modes of motion.
  Imaging is carried out with an objective placed at a working distance of 18 mm from the ions.
  }
  \label{fig:example_trap}
\end{figure*}

As an example of a possible implementation of a two-dimensional Penning trap array suitable for quantum simulation we consider a honeycomb lattice of 62 sites.
The surface-electrode pattern required for such a layout with nearest-neighbor spacing of 30 \micron~is shown in figure \ref{fig:example_trap}.
Applying a voltage of 135~V to the DC electrodes results in an axial trap frequency of $\omega_z = 2 \pi \cdot 2.1~\text{MHz}$ for beryllium ions.
In a global magnetic field angled at $\Theta = 20^\circ$ from the plane normal with a magnitude of $B_0 = 2.5$ T, we get a reduced cyclotron frequency of $\omega_+ = 2\pi \cdot 3.73~\text{MHz}$ and a magnetron frequency of $\omega_- = 2\pi \cdot 0.54~\text{MHz}$.
In this setup, effective laser cooling requires an axialization field, which is generated by applying a drive to the r.f.~electrodes at a frequency of $\omega_+ + \omega_- = 2\pi \cdot 4.27~\text{MHz}$ and an amplitude of 4 V.

Working inside the bore of a several tesla superconducting magnetic presents challenges for the delivery of the optical power required for cooling, detection, and single- and multi-qubit control operations.
One approach which has been used in multiple experiments is to direct beams down the magnet bore, and then re-direct them using mirrors through a vacuum system housed in the bore \cite{Biercuk:2009, Mavadia:NatComm:2013, Ball:RevSciInst:2019}.
For the purposes of working with ions in an array of surface traps with the ions at a height of 45 \micron~from the surface, we envision beams with a diameter of 2 mm directed down the bore of the magnet, which are then reflected towards the trap with a mirror, and then focused to a beam waist of 10 \micron~using a 100 mm focal length cylindrical lens.
This configuration results in an intensity variation of only 2.5\% over the trap area, where 100 \textmu W of 313 nm cooling light would have an intensity 8 times that of the saturation intensity, which would be sufficient for laser cooling.
Raman beams with a power of 4.3 mW can be directed parallel to the trap surface with the difference wave-vector chosen to address either the axial or the radial modes of motion.
This results in a one-photon Rabi frequency of $\Omega = 2\pi \cdot 300~\text{MHz}$, and for a detuning of 150 GHz a two-photon Rabi frequency of $\Omega_2 = 2\pi \cdot 300~\text{kHz}$.
% These are the reference values used in sections \ref{sec:spinspin} and \ref{sec:Qcomp} when quantifying laser-ion interaction strength for for quantum simulations and computation.

The final element required for running such a setup is the imaging system.
Here we envision using a Schwarzschild objective that is diffraction limited over a 130~\micron$^2$ region, with an effective focal length of 9.5 mm, a working distance of 18 mm, and a collection efficiency of 4\%, similar to what we have developed for previous experiments \cite{ThLeupold}.
Light collected from this objective is directed back out of the magnet bore with a mirror where it can be imaged onto a suitable detector.
Taking into account other losses and the quantum efficiency of the detector, we would expect a detection efficiency of 1\%, or 50 counts in 100 \textmu s.

It is worth considering the power dissipation due to the axialization field.
For a surface electrode trap fabricated using a Complementary-Metal-on-Silicon (CMOS) process, each ion's trap site (with electrode area of 1200 \micron$^2$) would be expected to have a capacitance of 0.01 pF.
Assuming a resistance of $R = 4$ ohms of the wiring to the trap chip this means that each trap site would dissipate $P_\text{RF} = \frac{1}{2} R C^2 V_\text{RF}^2 \Omega_\text{RF}^2 = 2.3~\text{pW}$ due to the axialization.
Such a treatment makes clear the power-dissipation advantage of the Penning trap approach.
For a Paul trap of similar dimensions and trap frequencies the required r.f.~voltage is $V_\text{RF} = 100 \text{\,V}$ at a drive frequency $\Omega_\text{RF} = 2\pi \times 163~\text{MHz}$ \cite{Wilson:Nature:2014}.
This would dissipate a power of  $P_\text{RF} = 2~\text{\textmu W}$, a factor of $10^6$ more.

\section{Scaling}

%%% FIX ME
Arrays of microfabricated Penning traps provide a new perspective towards scaling, because they do not require applying r.f.~drives to the ion trap chip beyond the modest frequencies and voltages needed for axialization.
The reduction in power-dissipation becomes even more critical for a larger number of sites, for instance an array similar to that considered above with thousands of sites (as might be required for quantum computation) would require dissipating several hundred watts in a similarly sized r.f.~trap, which is most likely not feasible.
By contrast, the Penning trap approach dissipates several hundred microwatts due to axialization fields, which only have to be applied during the reduced fraction of time devoted to Doppler cooling.

Penning traps also reduce the number of degrees of freedom which must be controlled at any one site relative to radio-frequency traps.
Stray electric fields in Paul traps lead to misalignment of the radio-frequency and static quadrupole potentials, resulting in undesirable micromotion which affects the interaction of the ions with laser fields.
In Penning traps, these move the center of the trap, resulting in shifts of the motional frequencies if the potential is anharmonic, but do not produce any other undesirable effect.
Recent evidence suggests that heating may be linked to processes driven by the radio-frequency drive in Paul traps \cite{17Hite}, which gives hope that anomalous heating could be reduced in the systems proposed here.

While the prospects for scaling look attractive relative to current approaches, many caveats remain.
Although co-wiring of thousands of electrodes is well within the capabilities of fabrication based on commercial CMOS processes, and trap chips fabricated using these methods have been successfully used to trap ions \cite{14Mehta}, much work on fabrication and operation remains in order to realize large-scale arrays.
In common with all ion trapping schemes requiring local control for quantum computing, a considerable remaining challenge is optical delivery of focused laser beams, in particular where local fields for individual sites are required (this constraint is not met in the approach considered in section \ref{sec:example-implementation} above, which uses global laser beams).
For truly large-scale quantum computing systems the integration of optics \cite{16Mehta, 20Mehta, 20Niffenegger} into the ion trap chips seems to be essential.
The need for high-intensity laser beam delivery might be mitigated using oscillating magnetic fields delivered directly from the ion trap chip, or using static magnetic field gradients \cite{19Srinivas, 16Harty}.

As opposed to Paul traps, the primary difficulty of realizing high-quality qubit control in Penning traps is the presence of a high magnetic field, fluctuations in which pose a limit to spin-state coherence.
Stability on the level of a part-per-billion is available through the use of superconducting magnets, which limits current experiments to coherence times of 50~ms \cite{Britton:2016aa}.
The effect on the stability of motional modes is negligible at this level.
One approach to protecting spin qubits in a quantum computation setting would be to use decoherence-free-subspaces, which protect against the expected homogeneous field fluctuations \cite{haffner2005robust}.
Alternative schemes might utilize dynamical decoupling or employ nuclear transitions (which are less sensitive to field fluctuations) for memory \cite{Biercuk09}.

\section{Conclusions}
This study establishes the possibility of using ions in Penning trap arrays for scalable many-body quantum simulations and quantum computation.
While we consider here simple settings for motional modes and the possibility to realize tunable-range spin-spin interactions, the flexibility of local control of trapping potentials means that couplings could be used to access a wide range of possibilities which have been previously discussed in the context of other systems, including but not limited to spin-boson systems \cite{08Porras}, dissipative simulations \cite{Barreiro2011}, and engineered topology \cite{Bermudez2011}.
Although quantum computing seems feasible using static arrays with selected ions tuned into local resonance, this is only one way of scaling  trapped-ion quantum information.
Breaking the lattice down into smaller units would allow smaller ion separations to be achieved \cite{16Mielenz}.
Utilizing this would require some level of transport of ions, which could be performed by moving the electric quadrupole positions dynamically \cite{98Wineland2,02Kielpinski, Hellwig2010, Crick2010}.
Here the Penning microtrap array holds the considerable advantage that there is no need to separate regions for quantum gates from specialized junction regions where 2-dimensional transport  occurs \cite{09Blakestad,14Shu}.
Since the homogeneous magnetic field supplies 3-dimensional confinement anywhere that a static quadrupole can be placed, re-organization of the potential landscape in 3 dimensions would allow 3-dimensional movement of ions at any point above the trap surface.
Thus Penning microtrap arrays appear to remove multiple constraints on scaling trapped-ion quantum computing, paving the way to useful quantum computers.

\bibliography{myrefs}

\appendix
\section{Single site Penning trap}\label{app:SingleSite}
We consider for the moment a simplified setting, in which each ion is trapped in a symmetric static quadrupole potential $m \omega_z^2/2 (z^2 - (x^2 + y^2)/2)$ for ions of mass $m$ and charge $e$ embedded in a magnetic field of strength $B_0$ aligned along the $z$ axis. At a single site, the potential and magnetic field give rise to a Hamiltonian
\begin{equation}
\begin{aligned}
\hat{H}_s &= \frac{\hat{p}^2_x + \hat{p}^2_y}{2 m} + \frac{1}{8} m \omega^2_1 \left( \hat{x}^2 + \hat{y}^2 \right) - \frac{\omega_c}{2}  \left( \hat{x} \hat{p}_y - \hat{y} \hat{p}_x \right) \\
&+ \frac{\hat{p}^2_z}{2 m} + \frac{1}{2} m \omega^2_z \hat{z}^2
\end{aligned}
\end{equation}
where $\omega_1 = \sqrt{\omega_c^2 - 2 \omega_z^2}$ and $\omega_c = e B_0/m$ is the bare cyclotron frequency. Writing the position and momentum operators in terms of creation and annihilation operators for the individual $x$, $y$, and $z$ degrees of freedom, defined as
\begin{equation}
\begin{aligned}
\hat{x} = \sqrt{\frac{ \hbar}{m \omega_1}} \left( \hat{a}^{\dagger}_x + \hat{a}_x \right) &, \quad \hat{p}_x = i \sqrt{\frac{\hbar m \omega_1}{4}} \left( \hat{a}^{\dagger}_x - \hat{a}_x \right), \\
\hat{y} = \sqrt{\frac{ \hbar}{m \omega_1}} \left( \hat{a}^{\dagger}_y + \hat{a}_y \right) &, \quad \hat{p}_y = i \sqrt{\frac{\hbar m \omega_1}{4}} \left( \hat{a}^{\dagger}_y - \hat{a}_y \right), \\
\hat{z}  = \sqrt{\frac{\hbar}{2 m \omega_z}} \left( \hat{a}^{\dagger}_z + \hat{a}_z \right) &, \quad \hat{p}_z = i \sqrt{\frac{\hbar m \omega_z}{2}} \left( \hat{a}^{\dagger}_z - \hat{a}_z \right), \\
\end{aligned}
\end{equation}
the Hamiltonian can be re-written as
\be
\hat{H}_s &=&
\frac{\hbar \omega_1}{2} \left(\hat{a}_x^\dagger \hat{a}_x+\hat{a}_y^\dagger \hat{a}_y + 1\right) + i \frac{\hbar \omega_c}{2}\left(\hat{a}_x^\dagger \hat{a}_y - \hat{a}_y^\dagger \hat{a}_x \right) \nonumber \\ && + \hbar \omega_z \left(\hat{a}_z^\dagger \hat{a}_z + 1/2\right) .
\ee
This can be separated into a sum of three independent harmonic oscillators using the transformation $
\hat{a}_{\pm} = \frac{1}{\sqrt{2}} \left( \hat{a}_x \pm i \hat{a}_y \right)$
for the radial motion. We then obtain the expression given in equation \ref{eq:HsinglePenning} in the main manuscript.

\section{Normal modes: classical description}\label{app:Classical}
We consider a system of $N$ micro-Penning traps containing a single ion each (with charge $+e$) arranged arbitrarily in space. The Coulomb interaction between ions leads to a coupling between their motional states, resulting in $3N$ collective normal modes of motion.
\\
\subsection*{Lagrangian Formulation}
Let the quadrupole center $j$ and the position of the ion $j$ in the reference frame of the lab be defined by the coordinates $\mathbf{D}_{j}$ and $\mathbf{R}_{j}$ respectively. Then the local coordinates of the ion $j$ with respect to this quadrupole center are given by the vector $\bar{\mathbf{r}}_{j}=\mathbf{R}_{j}-\mathbf{D}_{j}$. \\
The trapping electrodes create a static quadrupole electric potential centered at each site $j$ and this potential can be written in terms of the local coordinates as $\phi_j=\sum_{\mu\nu}\phi_{j0}^{\mu\nu}\bar{r}_{j}^{\mu}\bar{r}_j^{\nu}$, where the indices $\mu$ and $\nu$ run over the Cartesian components, $x$, $y$ and $z$.\\
The electrostatic potential acting on the ion $j$ due to the Coulomb interaction with other ions is
\begin{equation}
\kappa_j = \sum_{k\neq j}\frac{e}{4\pi\epsilon_0|\mathbf{R}_{j}-\mathbf{R}_{k}|} = k_e e \sum_{k\neq j}\frac{1}{|\mathbf{R}_{jk}|},
\end{equation}
where $k_e=1/(4\pi\epsilon_0)$ is the Coulomb constant.\\
The total electric potential, in the absence of any oscillating fields, is thus given by $\Phi_j = \phi_j + \kappa_j$.\\
A static homogeneous magnetic field $\mathbf{B}=B_0\sin\theta\cos\varphi\,\hat{\bf e}_x+B_0\sin\theta\sin\varphi\,\hat{\bf e}_y+B_0\cos\theta\,\hat{\bf e}_z$ creates the vector potential $\mathbf{A}_j$ at the site $j$. In the symmetric gauge, $\mathbf{A}_j=\frac{1}{2}(\mathbf{B}\times\mathbf{R}_j). $\\
In the laboratory frame of reference, the total Lagrangian of the system is then given by
\begin{equation}
L_{\text{tot}} = \sum^{N}_{j=1}\bigg\{ \frac{1}{2}m_j |\dot{\mathbf{R}}_{j}|^{2}+e\mathbf{A}_j \cdot \dot{\mathbf{R}}_j-e\Phi_{j}\bigg\},
\end{equation}
where $m_j$ is the mass of the $j$th ion.\\
The normal mode analysis begins by finding the equilibrium configuration of ions $\{ \mathbf{R}_{j0} \}$, which is determined by the minimum of the total potential energy. By expanding the system Lagrangian about the equilibrium position of each ion in a Taylor series up to second order, we get a Lagrangian in terms of the generalized position vectors $\mathbf{r}_j = \mathbf{R}_{j}-\mathbf{R}_{j0}$ which specify the displacement of each ion from its equilibrium point. The second order term in the expansion effectively determines the normal mode dynamics of the system near the stable spatial configuration and is given by

\begin{widetext}

\begin{equation}
\begin{aligned}
L &=\sum^{N}_{j=1}\bigg\{ \frac{1}{2}m_{j}|\dot{\mathbf{r}}_{j}|^{2}+ \frac{e}{2}(\mathbf{B}\times\mathbf{r}_j) \cdot \dot{\mathbf{r}}_j -e\sum_{\mu\nu}\phi_{j0}^{\mu\nu}r_{j}^{\mu}r_j^{\nu}  \bigg\}\\
&-\frac{k_e e^2}{2}\sum^{N}_{j=1}\sum^N_{k\neq j}\bigg\{ \sum_{\mu} \frac{3R^{\mu 2}_{jk0}-R^2_{jk0}}{R^5_{jk0}} (r^{\mu}_{j}-r^{\mu}_{k})^2 + \sum_{\mu \neq \nu}\frac{3R^{\mu}_{jk0}R^{\nu}_{jk0}}{R^5_{jk0}}(r^{\mu}_{j}-r^{\mu}_{k})(r^{\nu}_{j}-r^{\nu}_{k}) \bigg \}.
\end{aligned}
\end{equation}

\end{widetext}

We proceed by putting together all the generalized position coordinates into a single $3N$-dimensional vector $q=\begin{bmatrix} x_1 \, ...\, x_N & y_1 \,... \, y_N & z_1 \,... \, z_N\end{bmatrix}^T$ and introducing the
$3N\times3N$ block matrices $M$, $W$, $V$ and $K$ constructed in terms of $N\times N$ sub-matrices as

\begin{widetext}

\begin{equation}
\begin{aligned}
M=\begin{bmatrix} M^{xx} & \mathbb{O}_{N} & \mathbb{O}_{N} \\ \mathbb{O}_{N} & M^{yy} & \mathbb{O}_{N}  \\ \mathbb{O}_{N} & \mathbb{O}_{N} & M^{zz} \end{bmatrix} \qquad &, \qquad
W=eB_0\begin{bmatrix} \mathbb{O}_{N} & \cos\theta \cdot \mathbb{I}_{N}  & -\sin\theta\sin\varphi\cdot\mathbb{I}_{N} \\ -\cos\theta \cdot\mathbb{I}_{N} & \mathbb{O}_{N}  & \sin\theta\cos\varphi\cdot\mathbb{I}_{N} \\  \sin\theta\sin\varphi\cdot\mathbb{I}_{N} & -\sin\theta\cos\varphi\cdot\mathbb{I}_{N}  & \mathbb{O}_{N} \end{bmatrix},
\\
V=\begin{bmatrix} V^{xx} & V^{xy} & V^{xz} \\ V^{yx} & V^{yy} & V^{yz}  \\ V^{zx} & V^{zy} & V^{zz} \end{bmatrix} \qquad &, \qquad
K=\begin{bmatrix} K^{xx} & K^{xy} & K^{xz} \\ K^{yx} & K^{yy} & K^{yz}  \\ K^{zx} & K^{zy} & K^{zz} \end{bmatrix}.
\end{aligned}
\end{equation}

\end{widetext}

Here $\mathbb{I}_N$ and $\mathbb{O}_N$ represent the $N\times N$ identity and zero matrices respectively and the components of other sub-matrices are defined as
\begin{equation}
M^{\mu\mu}_{jk}=m_{j}\delta_{jk}
\end{equation}
\begin{equation}
V^{\mu\nu}_{jk}=2e\phi^{\mu \nu}_{j0}\delta_{jk},
\end{equation}
\begin{subequations}
\begin{equation}
K^{\mu\mu}_{jk}=
\begin{cases}
-k_e e^2\sum_{l\neq j}\frac{R^2_{jl0}-3R^{\mu2}_{jl0}}{R^5_{jl0}} &,j=k \\
k_e e^2\frac{R^2_{jk0}-3R^{\mu 2}_{jk0}}{R^5_{jk0}} &, j\neq k
\end{cases},
\end{equation}
\begin{equation}
K^{\mu \nu}_{jk}=
\begin{cases}
3k_e e^2\sum_{l\neq j}\frac{R^{\mu}_{jl0}R^{\nu}_{jl0}}{R^5_{jl0}} &,j=k \\
-3k_e e^2\frac{R^{\mu}_{jk0}R^{\nu}_{jk0}}{R^5_{jk0}} &, j\neq k
\end{cases},\mu \neq \nu,
\end{equation}
\end{subequations}
where indices $j$ and $k$ run from $1$ to $N$ while again the indices $\mu$ and $\nu$ refer to the components $x$, $y$ and $z$.\\
The above definitions together with $\Phi = V + K$ allow us to write the effective phonon Lagrangian compactly as
\begin{equation}
\begin{aligned}
L & =\sum^{3N}_{j=1}\bigg\{ \frac{1}{2}M_{jj}\dot{q}_{j}^{2} - \frac{1}{2}\sum^{3N}_{k=1}W_{jk}\dot{q}_{j}q_k-\frac{1}{2}\sum^{3N}_{k=1}\Phi_{jk}q_{j}q_k \bigg\}.
\end{aligned}
\end{equation}
It should be clear that $M$ is a real diagonal matrix while $W$ is a real antisymmetric matrix. The matrix $V$ is traceless as a direct consequence of Laplace's equation, while the matrix $K$ is traceless because the Coulomb forces being internal forces in the system of ions pairwise cancel each other and the total sum equates to zero. $V$ and $K$ are also both real and symmetric. As a result $\Phi = V + K$ is a real symmetric traceless matrix. These properties are useful in determining certain characteristics of the resulting normal mode eigenfrequencies and eigenvectors of the system.
\subsection*{\textsc{Equations of Motion}}
Through the Euler-Lagrange equations,
\begin{equation}
\frac{d}{dt}\bigg\{ \frac{\partial L}{\partial \dot{q}_j}\bigg\} =\frac{\partial L}{\partial q_j},
\end{equation}
we can derive from the Lagrangian the equations of motion of our system. The two relevant derivatives are
\begin{equation}
\frac{\partial L}{\partial \dot{q}_j} =M_{jj}\dot{q}_{j}-\frac{1}{2}\sum^{3N}_{k=1}W_{jk}q_k
\end{equation}
and
\begin{equation}
\frac{\partial L}{\partial q_j} =\frac{1}{2}\sum^{3N}_{k=1}W_{jk}\dot{q}_k -\sum^{3N}_{k=1}\Phi_{jk}q_k,
\end{equation}
which we can combine to get
%\begin{equation}
%M_{jj}\ddot{q}_{j}-\frac{1}{2}\sum^{3N}_{k=1}W_{jk}\dot{q}_k=\frac{1}{2}\sum^{3N}_{k=1}W_{jk}\dot{q}_k-\sum^{3N}_{k=1}\Phi_{jk}q_k
%\end{equation}
%or
\begin{equation}
M_{jj}\ddot{q}_{j}-\sum^{3N}_{k=1}W_{jk}\dot{q}_k+\sum^{3N}_{k=1}\Phi_{jk}q_k=0 \ .
\end{equation}
In vector form, we can then see that the equations of motion can be written as
\begin{equation}
\label{eq:qevequation}
M\ddot{q}-W\dot{q}+\Phi q=0 \ .
\end{equation}
To find the normal modes of motion, we substitute the oscillating trial solution $q=q_0 e^{-i\omega t}$ which yields the QEP
\begin{equation}\label{eq:QEPclassical}
[\,\omega^2(M +\omega (-iW)-\Phi\, ]q_0=0,
\end{equation}
that can be solved for complex eigenvectors $q_0$ and eigenvalues $\omega$, which in general can be complex. The set of eigenvalues $\{ \omega_{\lambda} \}$ are the normal mode frequencies while the corresponding normalized eigenvectors $\{q_{\lambda} \}$ give us the normal mode coordinates. \\
The general solution can be written as
\begin{equation}
q(t) = \sum^{3N}_{\lambda=1} \rho_{\lambda} q_{\lambda} e^{-i\omega_{\lambda}t},
\end{equation}
where $\rho_{\lambda}$ are complex scalars. The motion of the ions in terms of the normal modes can then be retrieved as
\begin{equation}
r(t) = \text{Re}(q(t)) = \frac{1}{2}\sum^{3N}_{\lambda=1} ( \rho_{\lambda} q_{\lambda} e^{-i\omega_{\lambda}t} + \rho^*_{\lambda} q^*_{\lambda} e^{i\omega_{\lambda}t}).
\end{equation}
For real frequencies, the collective motion is bounded and hence all ions are confined.\\

%--------------------

\section{Normal modes: quantum mechanical description}

From the Lagrangian of the system we can identify canonical conjugate variables to formulate our Hamiltonian. The generalized momentum corresponding to the generalized position $q_j$ is given by $
p_j=\frac{\partial L}{\partial \dot{q}_j} $. The classical Hamiltonian of the system is then
\begin{equation}
\begin{aligned}
H &=\sum^{3N}_{j=1}\dot{q}_j p_j - L\\
%&=\sum^{3N}_{j=1}\bigg\{ M_{jj}\dot{q}_{j}^{2}-\frac{1}{2}\sum^{3N}_{k=1}W_{jk}\dot{q}_{j}q_k\bigg\} - \sum^{3N}_{j=1}\bigg\{ \frac{1}{2}M_{jj}\dot{q}_{j}^{2}-\frac{1}{2}\sum^{3N}_{k=1}\Phi_{jk}q_{j}q_k - \frac{1}{2}\sum^{3N}_{k=1}W_{jk}\dot{q}_{j}q_k\bigg\}\\
&=\sum^{3N}_{j=1}\bigg\{\frac{1}{2} M_{jj}\dot{q}_{j}^{2}+\frac{1}{2}\sum^{3N}_{k=1}\Phi_{jk}q_{j}q_k \bigg\}.
\end{aligned}
\end{equation}
Quantizing the generalized coordinates as operators satisfying the standard commutation relations
\begin{equation}
[\hat{q}_j,\hat{q}_k]=0, \qquad [\hat{p}_j,\hat{p}_k]=0, \qquad [\hat{q}_j,\hat{p}_k]=i\hbar\delta_{jk},
\end{equation}
we can formulate the quantum mechanical Hamiltonian of the system as

\begin{widetext}

\begin{equation}
\begin{aligned}
\hat{H} =\sum^{3N}_{j=1}\bigg\{\frac{\hat{p}_{j}^{2}}{2M_{jj}}+\frac{1}{4M_{jj}}\sum^{3N}_{k=1}W_{jk}\hat{p}_{j}\hat{q}_k-\sum^{3N}_{k=1}\frac{W_{jk}}{4M_{kk}}\hat{q}_{j}\hat{p}_k -\frac{1}{8}\sum^{3N}_{k=1}T_{jk}\hat{q}_{j}\hat{q}_k+\frac{1}{2}\sum^{3N}_{k=1}\Phi_{jk}\hat{q}_{j}\hat{q}_k \bigg\}.
\end{aligned}
\end{equation}

\end{widetext}
where $T=WM^{-1}W$ is a real symmetric matrix.\\
To diagonalize the Hamiltonian in the second quantized form $\hat{H}=\sum^{3N}_{\lambda=1}\hbar\omega_{\lambda}(\hat{a}^{\dagger}_\lambda \hat{a}_{\lambda} + \frac{1}{2})$, we form the phonon creation and annihilation operators, $\hat{a}^{\dagger}_\lambda$ and $\hat{a}_{\lambda}$, for the mode $\lambda$ as linear combinations of the generalized position and momentum operators,
\begin{equation}
\hat{a}^{\dagger}_\lambda=\sum^{3N}_{k=1}(\alpha_{\lambda k}\hat{p}_k+\beta_{\lambda k}\hat{q}_k),
\end{equation}
\begin{equation}
\hat{a}_{\lambda}=\sum^{3N}_{k=1}(\alpha^{*}_{\lambda k}\hat{p}_k+\beta^{*}_{\lambda k}\hat{q}_k),
\end{equation}
where $\alpha_{\lambda k}$ and $\beta_{\lambda k}$ are complex numbers. For the commutation relation $[\hat{a}_{\lambda},\hat{a}^{\dagger}_{\lambda^{\prime}}]=\delta_{\lambda \lambda^{\prime}}$ to hold, the Hamiltonian must satisfy the commutation relation
\begin{equation}
[\hat{H},\hat{a}^{\dagger}_\lambda]=\hbar\omega_{\lambda} \hat{a}^{\dagger}_\lambda.
\end{equation}
This commutator can be calculated by substituting $\hat{H}$ and $\hat{a}^{\dagger}_\lambda$ in terms of the canonical variables and comparing the coefficients of $\hat{p}_l$ and $\hat{q}_l$ in $[\hat{H},\hat{a}^{\dagger}_\lambda]=\hbar\omega_{\lambda} \sum^{3N}_{l=1}(\alpha_{\lambda l}\hat{p}_l+\beta_{\lambda l}\hat{q}_l)$ yields the following set of coupled equations
\begin{subequations}
\begin{equation}
\begin{aligned}
-i\frac{\beta_{\lambda l}}{M_{ll}}+\frac{i}{2}\sum^{N}_{m=1}\frac{W_{lm}}{M_{ll}}\alpha_{\lambda m}
=\omega_{\lambda} \alpha_{\lambda l},
\end{aligned}
\end{equation}
\begin{equation}
\begin{aligned}
i \sum^{N}_{m=1} \bigg\{ \frac{W_{lm}}{2M_{mm}}\beta_{\lambda m}
-\frac{T_{lm}}{4}\alpha_{\lambda m}
+\Phi_{lm}\alpha_{\lambda m} \bigg\}
=\omega_{\lambda} \beta_{\lambda l}.
\end{aligned}
\end{equation}
\end{subequations}
These can be written more succinctly in vector form as
\begin{subequations}
\begin{equation}
\begin{aligned}
-i M^{-1}\beta_{\lambda}+\frac{i}{2}M^{-1}W\alpha_{\lambda}
=\omega_{\lambda} \alpha_{\lambda},
\end{aligned}
\end{equation}
\begin{equation}
\begin{aligned}
\frac{i}{2}WM^{-1}\beta_{\lambda}
-\frac{i}{4}T\alpha_{\lambda}
+i \Phi\alpha_{\lambda}
=\omega_{\lambda} \beta_{\lambda}.
\end{aligned}
\end{equation}
\end{subequations}
On eliminating $\beta_{\lambda}$ using $\beta_{\lambda}=i\omega_{\lambda} M\alpha_{\lambda}+\frac{1}{2}W\alpha_{\lambda}$, we then see that
%\begin{equation}
%\begin{aligned}
%\frac{i\hbar}{2}WM^{-1}(i\omega_{\lambda} M\alpha_{\lambda}+\frac{1}{2}W\alpha_{\lambda})
%-\frac{i\hbar}{4}T\alpha_{\lambda} +i\hbar \Phi\alpha_{\lambda}
%=\hbar\omega_{\lambda} (i\omega_{\lambda} M\alpha_{\lambda}+\frac{1}{2}W\alpha_{\lambda}),
%\end{aligned}
%\end{equation}
%or
%\begin{equation}
%\begin{aligned}
%\frac{i\hbar}{4}T\alpha_{\lambda}-\frac{\hbar}{2}\omega_{\lambda} W\alpha_{\lambda}
%-\frac{i\hbar}{4}T\alpha_{\lambda}
%+i\hbar \Phi\alpha_{\lambda}
%=i\hbar\omega^2_\lambda M \alpha_{\lambda}+\frac{\hbar}{2}\omega_{\lambda} W\alpha_{\lambda},
%\end{aligned}
%\end{equation}
%or
\begin{equation}
\begin{aligned}
\omega^2_\lambda M \alpha_{\lambda}-i\omega_{\lambda} W\alpha_{\lambda}-\Phi\alpha_{\lambda}=0,
\end{aligned}
\end{equation}
which is the same QEP encountered in the classical analysis in Appendix \ref{app:Classical}.
The QEP yields $6N$ eigenvectors and $6N$ eigenvalues. $3N$ eigenvectors will be used to form the creation operators while the other $3N$ eigenvectors to form the annihilation operators. We note that if the pair $(\nu_{\lambda},u_{\lambda})$ satisfies the QEP then the pair $(-\nu_{\lambda},u^*_{\lambda})$ also satisfies the QEP. Thus the total set of $6N$ eigenpairs
\begin{equation}
\begin{aligned}
S = \{ (\nu_{\lambda},u_{\lambda}) , \quad \mid \quad \nu^2_\lambda M u_\lambda-i\nu_\lambda W u_\lambda-\Phi u_{\lambda}=0   \}
\end{aligned}
\end{equation}
for $\lambda$ running over $1$ to $6N$ can be divided into two equally sized subsets depending on the signs of the eigenvalues:
\begin{equation}
\begin{aligned}
S_+ := \{ (\nu_{\lambda},u_{\lambda}) \quad &\mid \quad (\nu_{\lambda},u_{\lambda}) \in S, \quad \nu_{\lambda}> 0    \}, \\
S_- := \{ (-\nu_{\lambda},u^*_{\lambda}) \quad &\mid \quad (\nu_{\lambda},u_{\lambda}) \in S, \quad \nu_{\lambda}> 0    \}.
\end{aligned}
\end{equation}
The index $\lambda$ now runs from $1$ to $3N$ so that $\nu_{\lambda}$ is assumed to be positive from hereon.\\
Selecting the $3N$ eigenpairs which form the creation operators effectively means picking the sign of the eigenfrequency (and the corresponding eigenvector) for a given mode $\lambda$ in $\hat{H}=\sum^{3N}_{\lambda=1}\hbar\omega_{\lambda}(\hat{a}^{\dagger}_\lambda \hat{a}_{\lambda} + \frac{1}{2})$ and involves fixing the normalization of the eigenvectors $\alpha_{\lambda}$ so that $\left[ \hat{a}_{\lambda},\hat{a}^{\dagger}_{\lambda} \right]=1$. Explicitly,
\begin{equation}
\begin{aligned}
\left[ \hat{a}_{\lambda},\hat{a}^{\dagger}_{\lambda} \right] &=i\hbar ( \beta^{H}_{\lambda}\alpha_{\lambda}-\alpha^{H}_{\lambda }\beta_{\lambda})\\
&=\frac{\hbar}{\omega_{\lambda}}\left( \omega^2_{\lambda}\alpha^{H}_{\lambda}M\alpha_{\lambda}+\alpha^{H}_{\lambda}\Phi \alpha_{\lambda} \right).
\end{aligned}
\end{equation}
Substituting $\alpha_{\lambda}=c_\lambda\gamma_{\lambda}$, where $\gamma_{\lambda}$ is normalized to one and $c_{\lambda}$ is a complex scalar,
\begin{equation}
\begin{aligned}
\left[ \hat{a}_{\lambda},\hat{a}^{\dagger}_{\lambda} \right]&=\frac{\hbar |c_\lambda|^2}{\omega_{\lambda}}\big\{\omega^2_{\lambda}\gamma^{H}_{\lambda}M\gamma_{\lambda}+\gamma^{H}_{\lambda}\Phi \gamma_{\lambda} \big\},
\end{aligned}
\end{equation}
which for the condition $\left[ \hat{a}_{\lambda},\hat{a}^{\dagger}_{\lambda} \right] = 1$ yields
\begin{equation}
\begin{aligned}
|c_\lambda|^2=\frac{\omega_{\lambda}}{\hbar}\bigg\{ \frac{1}{\omega^2_{\lambda}\gamma^{H}_{\lambda}M\gamma_{\lambda}+\gamma^{H}_{\lambda}\Phi \gamma_{\lambda}} \bigg\}.
\end{aligned}
\end{equation}
Since $|c_\lambda|^2$ is non-negative, $\left[ \hat{a}_{\lambda},\hat{a}^{\dagger}_{\lambda} \right]=1$ only when the quantity $\omega_{\lambda}/(\omega^2_{\lambda}\gamma^{H}_{\lambda}M\gamma_{\lambda}+\gamma^{H}_{\lambda}\Phi \gamma_{\lambda})$ is positive. This expression helps us pick out the $3N$ eigenpairs to form the creation operators
\begin{equation}
(\omega_{\lambda},\alpha_{\lambda}) =
\begin{cases}
(\nu_{\lambda},c_{\lambda} \gamma_{\lambda}) \quad &,  \nu^2_{\lambda}\gamma^{H}_{\lambda}M\gamma_{\lambda}+\gamma^{H}_{\lambda}\Phi \gamma_{\lambda} > 0  \\
(-\nu_{\lambda},c_{\lambda}\gamma^*_{\lambda}) \quad &,  \nu^2_{\lambda}\gamma^{H}_{\lambda}M\gamma_{\lambda}+\gamma^{H}_{\lambda}\Phi \gamma_{\lambda} < 0  \\
\end{cases},
\end{equation}
where
\begin{equation}
\begin{aligned}
c_{\lambda}=\sqrt[]{\frac{\nu_{\lambda}}{\hbar|\nu^{2}_{\lambda}\gamma^H_{\lambda}M\gamma_{\lambda}+\gamma^H_{\lambda}\Phi \gamma_{\lambda}|}}
\end{aligned}
\end{equation}
can without loss of generality be chosen as real. Further, inverting the expressions for the creation and annihilation operators yields the second quantized form of the position and momentum operators,
\begin{equation}
\begin{aligned}
\hat{q}_j&=-i\hbar\sum^{3N}_{\lambda=1}(\alpha^{*}_{\lambda j}\hat{a}^{\dagger}_{\lambda}-\alpha_{\lambda j} \hat{a}_{\lambda})\\
&=-i\hbar\sum^{3N}_{\lambda=1}c_{\lambda}(\gamma^{*}_{\lambda j}\hat{a}^{\dagger}_{\lambda}-\gamma_{\lambda j} \hat{a}_{\lambda})
\end{aligned}
\end{equation}
and
\begin{equation}
\begin{aligned}
\hat{p}_j&=i\hbar\sum^{3N}_{\lambda=1}(\beta^{*}_{\lambda j}\hat{a}^{\dagger}_{\lambda}-\beta_{\lambda j} \hat{a}_{\lambda})\\
\end{aligned}.
\end{equation}

%-----------------
\section{Trap imperfections} \label{sec:AppInvarianceMass}
In a real experimental setup, the trapping potential may not be of the idealized form expected from the optimization of the electrode structures, while the magnetic field could be misaligned with the confining direction of the potential. As long as the imperfections in the electric potential are harmonic and the magnetic field is homogeneous over the entire system, we could employ the general discussion in Appendix \ref{app:Classical} in order to study the normal modes of the imperfect system. \\
Linearization of the QEP \ref{eq:QEPclassical} in the first-companion form yields the GEP
\begin{equation}
\begin{bmatrix} \mathbb{O}_{3N} & \mathbb{I}_{3N} \\ \Phi & iW \end{bmatrix} \begin{bmatrix} q_0\\ \omega q_0\end{bmatrix}
= \omega \begin{bmatrix} \mathbb{I}_{3N} &  \mathbb{O}_{3N} \\  \mathbb{O}_{3N} & M\end{bmatrix}\begin{bmatrix} q_0\\ \omega q_0\end{bmatrix}.
\end{equation}
Inversion of the matrix on the r.h.s. of the above equation leads to a further reduction to the SEP
\begin{equation}\label{eq:SEP}
A v = \omega v,
\end{equation}
with $6N$-dimensional eigenvectors $ v = \begin{bmatrix} q_0 & \omega q_0\end{bmatrix}^{T}$ and $6N$ eigenvalues $\omega$ belonging to the $6N \times 6N$ matrix
\begin{equation}
A= \begin{bmatrix} \mathbb{I}_{3N} &  \mathbb{O}_{3N} \\  \mathbb{O}_{3N} & M^{-1}\end{bmatrix}  \begin{bmatrix} \mathbb{O}_{3N} & \mathbb{I}_{3N} \\ \Phi & iW \end{bmatrix}.
\end{equation}
For the sake of brevity, we define the matrices $W^{\prime} = M^{-1} W$ and $\Phi^{\prime} = M^{-1} \Phi$, so that we have
\begin{equation}
A=   \begin{bmatrix} \mathbb{O}_{3N} & \mathbb{I}_{3N} \\ \Phi^{\prime} & iW^{\prime} \end{bmatrix}
\end{equation}
and
\begin{equation}
A^2=\begin{bmatrix} \Phi^{\prime} & iW^{\prime} \\ iW^{\prime}\Phi' & \Phi'-W^{\prime 2} \end{bmatrix}.
\end{equation}
Since $A^2 v =\omega^2 v$ and the sum of eigenvalues of a matrix is equal to its trace,
\begin{equation}
\begin{aligned}
\sum^{6N}_{\lambda=1}\omega^2_\lambda = \text{tr}(A^2)
= \text{tr}(2\Phi^{\prime}-W^{\prime 2})
=-\text{tr}(W^{\prime 2}),
\end{aligned}
\end{equation}
where we use the fact that $\Phi^{\prime}$ is traceless. The stability of the system can as usual be determined by checking if all eigenvalues are real. Noting that the frequencies come in pairs of positive and negative values in the stable regime we can express the above sum in terms of the $3N$ positive frequencies,
\begin{equation}
\begin{aligned}
\sum^{3N}_{\lambda=1}\omega^2_\lambda =-\frac{1}{2}\text{tr}(W^{\prime 2}).
\end{aligned}
\end{equation}
This trace can be explicitly calculated as
\begin{equation}
\begin{aligned}
\text{tr}(W^{\prime 2})&=-2 e^2 B^2_0 \sum^{N}_{j=1} \frac{1}{m^2_j}\\
&= -2 \sum^{N}_{j=1} \omega^2_{c,j},
\end{aligned}
\end{equation}
where $\omega_{c,j }= eB_0/m_j$ is the true cyclotron frequency of the $j$th ion, thus allowing us to express the sum as
\begin{equation}
\begin{aligned}
\sum^{3N}_{\lambda=1}\omega^2_\lambda =\sum^{N}_{j=1} \omega^2_{c,j}.
\end{aligned}
\end{equation}
Note that the sum is independent of the trapping potential. For a typical experiment with ions of the same species and no impurity defects, $m_j = m$, and the above sum further simplifies in terms of the common true cyclotron frequency $\omega_{c }= eB_0/m$ to
\begin{equation}\label{eq:GIT}
\begin{aligned}
\sum^{3N}_{\lambda=1}\omega^2_\lambda =N \omega^2_{c} .
\end{aligned}
\end{equation}
Equation \ref{eq:GIT} can be treated as a non-trivial generalization of the well known Brown-Gabrielse invariance theorem for a single ion in an imperfect Penning trap,
\begin{equation}
\omega^2_+ + \omega^2_- +\omega^2_z = \omega^2_c.
\end{equation}
One additional result can be derived from equation \ref{eq:SEP} by using the property that the product of the eigenvalues of a matrix is equal to its determinant so that
\begin{equation}
\begin{aligned}
\prod^{6N}_{\lambda=1}\omega_{\lambda} &= \lvert A \rvert \\
&= \begin{vmatrix} \mathbb{O}_{3N} & \mathbb{I}_{3N} \\ \Phi^{\prime} & iW^{\prime} \end{vmatrix}.
\end{aligned}
\end{equation}
An interchange of $3N$ columns in the matrix on the r.h.s. allows us to write the product in terms of the $3N$ positive frequencies as
\begin{equation}
\begin{aligned}
(-1)^{3N}\prod^{3N}_{\lambda=1}\omega^2_{\lambda} &= (-1)^{3N} \begin{vmatrix}  \mathbb{I}_{3N} & \mathbb{O}_{3N} \\  iW^{\prime} & \Phi^{\prime} \end{vmatrix}\end{aligned},
\end{equation}
or
\begin{equation}
\begin{aligned}
\prod^{3N}_{\lambda=1}\omega^2_{\lambda} &=  \lvert \Phi^{\prime} \rvert .
\end{aligned}
\end{equation}
Finally, we arrive at
\begin{equation}
\begin{aligned}
\prod^{3N}_{\lambda=1}\omega_{\lambda} &= \sqrt{ \lvert \Phi^{\prime} \rvert },
\end{aligned}
\end{equation}
which for the case of ions having identical masses can be more conveniently expressed as
\begin{equation}
\prod^{3N}_{\lambda=1} \left( m \omega^2_{\lambda} \right) =  \lvert \Phi \rvert .
\end{equation}
This result tells us that the product of eigenvalues is independent of the magnetic field and depends only on the curvature tensor of the total electric potential.

%--------------------------
\section{Spin spin coupling}\label{app:SpinSpin}

The derivation in this Appendix follows closely the methodology from ref. \cite{13Wang}. The ODF leads to the interaction term
\begin{equation}
\hat{H}_{\text{ODF}} = - \sum^N_{j=1} E_O \cos(\mathbf{k}_R\cdot\mathbf{R}_j - \mu_R t)\hat{\sigma}^z_j.
\end{equation}
In the Lamb-Dicke regime, we can expand this expression in terms of the equilibrium positions and deviations from them as
\begin{equation}
\hat{H}_{\text{ODF}} \approx  \sum^N_{j=1} E_O \mathbf{k}_R\cdot\hat{\mathbf{r}}_{j} \sin(\mathbf{k}_R\cdot\mathbf{R}_{j0} - \mu_R t)\hat{\sigma}^z_j.
\end{equation}
Then the effective spin Hamiltonian is given by
\begin{equation}
\label{eq:Hspin}
\hat{H}_{\text{SPIN}}= \frac{i}{2\hbar} [\hat{W}_I (t), \hat{V}_I (t)],
\end{equation}
which uses the definitions
\begin{equation}
\label{eq:VI}
\hat{V}_I(t) = e^{i \hat{H}_{\text{PH}}t/\hbar} \hat{H}_{\text{ODF}}(t)e^{-i \hat{H}_{\text{PH}}t/\hbar},
\end{equation}
\begin{equation}
\label{eq:WI}
\hat{W}_I (t) =\int_0^{t} \hat{V}_I(t') dt'
\end{equation}
and
\begin{equation}
\hat{H}_{\text{PH}}=\sum^{3N}_{\lambda=1}\hbar\omega_{\lambda}(\hat{a}^{\dagger}_{\lambda}\hat{a}_{\lambda} + \frac{1}{2}) \ .
\end{equation}
We can express the excursions from equilibrium in terms of the harmonic oscillator creation and annihilation operators, giving
\begin{equation}
\mathbf{k}_R\cdot\hat{\mathbf{r}}_{j} = -i\hbar \sum_{\nu} k^{\nu}_R \sum^{3N}_{\lambda=1} ( \alpha^*_{\lambda j \nu} \hat{a}^{\dagger}_{\lambda} - \alpha_{\lambda j \nu} \hat{a}_{\lambda}) \ .
\end{equation}
In the interaction picture with respect to the oscillator mode Hamiltonian $\hat{H}_{\text{PH}}$, we then find that
%\begin{equation}
%\hat{H}_{\text{ODF}} = -i\hbar E_O \sum^N_{j=1} \sin(\mathbf{k}_R\cdot\mathbf{R}_{j0} - \mu_R t)  \sum_{\nu} k^{\nu}_R \sum^{3N}_{\lambda=1} ( \alpha^*_{\lambda j \nu} \hat{a}^{\dagger}_{\lambda} - \alpha_{\lambda j \nu} \hat{a}_{\lambda}) \hat{\sigma}^z_j
%\end{equation}
%and
%\begin{equation}
%\hat{V}_I (t) = -i\hbar E_O \sum^N_{j=1} \sin(\mathbf{k}_R\cdot\mathbf{R}_{j0} - \mu_R t)  \sum_{\nu} k^{\nu}_R \sum^{3N}_{\lambda=1} ( \alpha^*_{\lambda j \nu} e^{i \hat{H}_{\text{PH}}t/\hbar} \hat{a}^{\dagger}_{\lambda}  \hat{\sigma}^z_j e^{-i \hat{H}_{\text{PH}}t/\hbar} - \alpha_{\lambda j \nu} e^{i \hat{H}_{\text{PH}}t/\hbar} \hat{a}_{\lambda}  \hat{\sigma}^z_j e^{-i \hat{H}_{\text{PH}}t/\hbar})
%\end{equation}
%Using the B-C-H formula,
%\begin{subequations}
%\begin{equation}
%e^{i \hat{H}_{\text{PH}}t/\hbar} \hat{a}^{\dagger}_{\lambda}  \hat{\sigma}^z_j e^{-i \hat{H}_{\text{PH}}t/\hbar} = e^{i \omega_{\lambda} t} \hat{a}^{\dagger}_{\lambda}  \hat{\sigma}^z_j
%\end{equation}
%\begin{equation}
%e^{i \hat{H}_{\text{PH}}t/\hbar} \hat{a}_{\lambda}  \hat{\sigma}^z_j e^{-i \hat{H}_{\text{PH}}t/\hbar} = e^{-i \omega_{\lambda} t} \hat{a}_{\lambda}  \hat{\sigma}^z_j
%\end{equation}
%\end{subequations}
%and hence we get
\begin{equation}
\begin{aligned}
\hat{V}_I (t)
% &= -i\hbar E_O \sum^N_{j=1} \sin(\mathbf{k}_R\cdot\mathbf{R}_{j0} - \mu_R t)  \sum_{\nu} k^{\nu}_R \sum^{3N}_{\lambda=1} ( \alpha^*_{\lambda j \nu} e^{i \omega_{\lambda} t} \hat{a}^{\dagger}_{\lambda}  \hat{\sigma}^z_j - \alpha_{\lambda j \nu} e^{-i \omega_{\lambda} t} \hat{a}_{\lambda}  \hat{\sigma}^z_j) \\
&=
-\frac{\hbar E_O}{2} \sum_{j,\nu,\lambda} k^{\nu}_R (f_{\lambda j }(t) \alpha^*_{\lambda j \nu} \hat{a}^{\dagger}_{\lambda}  \hat{\sigma}^z_j - g_{\lambda j }(t) \alpha_{\lambda j \nu} \hat{a}_{\lambda}  \hat{\sigma}^z_j),
\end{aligned}
\end{equation}
where we have defined the functions
\begin{subequations}
\begin{equation}
f_{\lambda j }(t) \equiv e^{i \phi_j}e^{i(\omega_{\lambda}-\mu_R) t} - e^{-i \phi_j}e^{i(\omega_{\lambda}+\mu_R) t},
\end{equation}
\begin{equation}
g_{\lambda j }(t) \equiv e^{i \phi_j}e^{-i(\omega_{\lambda}+\mu_R) t} - e^{-i \phi_j}e^{-i(\omega_{\lambda}-\mu_R) t},
\end{equation}
\begin{equation}
\phi_j = \mathbf{k}_R\cdot\mathbf{R}_{j0} \ .
\end{equation}
\end{subequations}
From equations \ref{eq:Hspin}, \ref{eq:VI} and \ref{eq:WI}, and making the rotating wave approximation with respect to the oscillator frequencies, we then find that %
the static part of the effective spin Hamiltonian can be written in the form of an Ising-like spin Hamiltonian
\begin{equation}
\hat{ H } _ {\text{SPIN}} = \sum _ { j j ^ { \prime } } J^0 _ { j j ^ { \prime } }  \hat{\sigma} _ { j } ^ { z } \hat{\sigma} _ { j ^ { \prime } } ^ { z },
\end{equation}
with the coupling terms given by
\begin{widetext}
\begin{equation}
\begin{aligned}
J^0 _ { j j ^ { \prime } }
&=\frac{E^2_O}{2}\sum_{\nu,\nu^{\prime}}\sum_{\lambda} \frac{\omega^2_{\lambda}}{m\omega^2_{\lambda} + \gamma^H_{\lambda}\Phi\gamma_{\lambda}} \frac{k^{\nu}_R k^{\nu^{\prime}}_R}{\mu^2_R -\omega^2_{\lambda}} \cos(\phi_j - \phi_{j^{\prime}}) \text{Re}(\gamma^*_{\lambda j \nu} \gamma_{\lambda j \nu^{\prime} }) \\
&- \frac{E^2_O}{2}\sum_{\nu,\nu^{\prime}}\sum_{\lambda} \frac{\omega_{\lambda} \mu_R}{m\omega^2_{\lambda} + \gamma^H_{\lambda}\Phi\gamma_{\lambda}} \frac{k^{\nu}_R k^{\nu^{\prime}}_R}{\mu^2_R -\omega^2_{\lambda}} \sin(\phi_j - \phi_{j^{\prime}}) \text{Im}(\gamma^*_{\lambda j \nu} \gamma_{\lambda j^{\prime} \nu^{\prime} }).
\end{aligned}
\end{equation}
\end{widetext}
Here $\gamma_{\lambda}$ is the normalized normal mode eigenvector corresponding to the frequency $\omega_{\lambda}$ and the indices $\nu,\nu'$ run over $x, y, z$.
%\end{document}

\section{Quantum gates}\label{app:Qgates}
The theory for calculations of quantum gates is similar to that detailed above, with the difference that for fidelity calculations we have to consider the effects of residual spin-motional entanglement and geometric phases from motional state components displaced in phase space. For a gate operated using an oscillating force with state-dependence in the $z$-basis, a common method for measuring the fidelity is to sandwich the two-ion gate in a Ramsey experiment performed simultaneously on both ions. This produces the maximally entangled state $\ket{\psi_{\rm B}} = \left(\ket{00} - i \ket{11}\right)/\sqrt{2}$. Thus the force pulse acts on a superposition state $\left(\ket{00} + \ket{01}+\ket{10}+\ket{11} \right)/2 \equiv \sum_{E  } \ket{E}/2$ with $E \in \{ 00,01,10,11 \}$. If the motional state of all modes is prepared in the ground state, the initial state of internal and motional states is
\be
\ket{\psi} = \frac{1}{2}\sum_{E} \ket{E} \otimes \bigotimes_\lambda \ket{0}_\lambda \ .
\ee
 
For two ions with index $j = 1$ and $j = 2$ which are at the same phase of the optical dipole force (here assumed to be zero), the interaction picture Hamiltonian $\hat{V}_I(t)$ becomes
\begin{equation}
\begin{aligned}
\hat{V}_I (t)
% &= -i\hbar E_O \sum^N_{j=1} \sin(\mathbf{k}_R\cdot\mathbf{R}_{j0} - \mu_R t)  \sum_{\nu} k^{\nu}_R \sum^{3N}_{\lambda=1} ( \alpha^*_{\lambda j \nu} e^{i \omega_{\lambda} t} \hat{a}^{\dagger}_{\lambda}  \hat{\sigma}^z_j - \alpha_{\lambda j \nu} e^{-i \omega_{\lambda} t} \hat{a}_{\lambda}  \hat{\sigma}^z_j) \\
&=
-\frac{\hbar E_O}{2} \sum_{\nu,\lambda} k^{\nu}_R \left[f_{\lambda}(t) \hat{a}^{\dagger}_{\lambda} \left(\alpha^*_{\lambda 1 \nu}  \hat{\sigma}^z_1+\alpha^*_{\lambda 2 \nu}  \hat{\sigma}^z_2\right) + {\rm h.c.} \right],
\end{aligned}
\end{equation}
where $\rm h.c.$ indicates the Hermitian conjugate. Since the gate acts on the four eigenstates of $\sigma^z_1 + \hat{\sigma}^z_2$, the action of the Hamiltonian acts in terms of these four eigenstates as
\be
U(t)\ket{\psi} = \frac{1}{2}\sum_{E} \ket{E} \prod_{\lambda}\hat{D}_\lambda(\chi_{\lambda, E}(t)) e^{i\Phi_{\lambda,E}(t)}
\ee
where $\hat{D}_\lambda$ denotes a displacement operator on the mode $\lambda$, with the displacement amplitude
\be
\chi_{\lambda, E}(t) = E_O p_{\lambda, E} F(\mu_R, \omega_\lambda, t) \ ,
\ee
with the function 
\be
F(\mu, \omega, t) &=& - \frac{e^{-i t (\mu - \omega)/2} \sin(t (\mu - \omega)/2)}{(\mu - \omega)} \nonumber \\ 
&+& \frac{e^{i t (\mu + \omega)/2} \sin(t (\mu + \omega)/2)}{(\mu + \omega)}
\ee
and $p_{\lambda, E} = \sum_\nu k_R^\nu \left(\pm \alpha_{\lambda 1 \nu} + \pm \alpha_{\lambda 2 \nu}\right)$ with plus signs for internal state $0$ and minus signs for internal state $1$ of each ion. The phases are given by
\be
\Phi_{\lambda, E}(t) = \frac{E_0^2}{4}\sum_{\nu, \nu'} |p_{\lambda, E}|^2 G(\mu_R, \omega_\lambda, t)
\ee
with
\be
G(\mu, \omega, t) &=& \frac{2 t \omega}{\mu^2 - \omega^2 } + \frac{2 \mu \sin( (\mu +\omega)t )}{(\mu + \omega)(\mu^2 - \omega^2)} \nonumber \\
&-& \frac{2 \mu \sin((\mu -\omega) t)}{(\mu - \omega)(\mu^2 - \omega^2)}  - \frac{\omega \sin(2\mu t)}{\mu (\mu^2 - \omega^2) } \ .
\ee
Once the state after the gate has been found, it can be used to form a density matrix from which all relevant quantities can be obtained. For the fidelity with the maximally entangled state, we take $F = \bra{\psi_B}\rho(t) \ket{\psi_B}$. We trace out the motional displacements using the overlap between two coherent states
\be
\bra{0} \hat{D}^\dagger(\chi) \hat{D}(\beta) \ket{0} = e^{-|\chi|^2/2}e^{-|\beta|^2/2}e^{\chi^*\beta} \ .
\ee

\end{document}